\documentclass[apj]{emulateapj}  
\usepackage{graphicx}
\usepackage{txfonts}
\usepackage{graphicx}
\begin{document}

\title{Extremely Metal-Poor Galaxies: The Environment}

\author{M. E. Filho\altaffilmark{1,2,3,4,5}, 
J. S\'anchez Almeida\altaffilmark{2,3},
C. Mu\~noz-Tu\~n\'on\altaffilmark{2,3},
S. E. Nuza\altaffilmark{6},
F. Kitaura\altaffilmark{6}
and
S. He{\ss}\altaffilmark{6}
}

\altaffiltext{1}{Universidad de Las Palmas de Gran Canaria-Universidad de La Laguna, CIE Canarias: Tri-Continental Atlantic Campus, Canary Islands, Spain}
\altaffiltext{2}{Instituto Astrof\'\i sica de Canarias, 38200 La Laguna, Tenerife, Spain} 
\altaffiltext{3}{Departamento de Astrof\'\i sica, Universidad La Laguna, 38206 La Laguna, Tenerife, Spain} 
\altaffiltext{4}{Centro de Astrof\'isica da Universidade do Porto, Rua das Estrelas s/n, 4150-762 Porto, Portugal}
\altaffiltext{5}{Instituto de Astrof\'\i sica e Ci\^encias do Espa\c co,  Universidade de Lisboa, OAL, Tapada da Ajuda, 1349-018 Lisboa, Portugal}
\altaffiltext{6}{Leibniz-Institut f\"ur Astrophysik Potsdam (AIP), An der Sternwarte 16, D-14482 Potsdam, Germany}

\email{email: mfilho@astro.up.pt}

%\date{}

\shortauthors{M. E. Filho et al.}
\shorttitle{The Environment of Extremely Metal-Poor Galaxies}

\begin{abstract}

We have analyzed bibliographical observational data and theoretical predictions, in order to probe the environment in which extremely metal-poor dwarf galaxies (XMPs) reside. We have assessed the \ion{H}{1} component and its relation to the optical galaxy, the cosmic web type (voids, sheets, filaments and knots), the overdensity parameter and analyzed the nearest galaxy neighbours. The aim is to understand the role of interactions and cosmological accretion flows in the XMP observational properties, particularly the triggering and feeding of the star formation. We find that XMPs behave similarly to Blue Compact Dwarfs; they preferably populate low-density environments in the local Universe: $\sim$ 60\% occupy underdense regions, and $\sim$ 75\% reside in voids and sheets. This is more extreme than the distribution of irregular galaxies, and in contrast to those regions preferred by elliptical galaxies (knots and filaments). We further find results consistent with previous observations; while the environment does determine the fraction of a certain galaxy type, it does not determine the overall observational properties. With the exception of five documented cases (four sources with companions and one recent merger), XMPs do not generally show signatures of major mergers and interactions; we find only one XMP with a companion galaxy within a distance of 100~kpc, and the \ion{H}{1} gas in XMPs is typically well-behaved, demonstrating asymmetries mostly in the outskirts. We conclude that metal-poor accretion flows may be driving the XMP evolution. Such cosmological accretion could explain all the major XMP observational properties: isolation, lack of interaction/merger signatures, asymmetric optical morphology, large amounts of unsettled, metal-poor \ion{H}{1} gas, metallicity inhomogeneities, and large specific star formation.

\end{abstract}

\keywords {galaxies: dwarf -- galaxies: formation -- galaxies: evolution -- galaxies: interaction -- galaxies: starburst}

\maketitle

\section{Introduction}

%J1647+2105 SF knot in the tidal tail of a merger and J0014+0044 HII region in UGC139

The standard model of galaxy formation and evolution includes hierarchical merging/clustering, where smaller structures are used as building blocks of more massive galaxies, and evolution through accretion from cosmological gas filaments in the cosmic web (e.g.,~Combes 2004; Dekel et al. 2009; Dekel, Sari \& Ceverino 2009; Nuza et al. 2014a; S\'anchez Almeida et al. 2014b). Because merger events are generally responsible for the disruption of gas and stellar aggregates, they have been traditionally associated with the triggering of supermassive black hole accretion at the center of galaxies (Active Galactic Nuclei activity; Springel et al. 2005; Hopkins et al. 2008; Di Matteo et al. 2008), and with the generation of star formation episodes (e.g.,~Sanders et al. 1988; Robaina et al. 2009). However, the contribution of cosmological cold-accretion flows to galaxy evolution in general, and star formation in particular, is recently gaining ground. Indeed, cosmological accretion onto galaxies of external metal-poor gas is predicted by numerical simulations to be the main mode of galaxy and disk assembly, at high, and possibly also at low, redshift (Birnboim \& Dekel 2003; Brooks et al. 2009). The theory predicts that these cosmological filaments are the main process by which star formation is fed and triggered at all cosmological times (Brooks et al. 2009; Verbeke et al. 2014). Cosmological gas accretion should have been more common at high redshift, but in the local Universe, it is hypothesized to intervene in the evolution of individual low-mass galaxies in low-density environments, via small-scale gas streams, facilitating the arrival of gas from large distances (Kere\v{s} et al. 2005; Dekel \& Birnboim 2006; Dekel et al. 2009; Dekel, Sari \& Ceverino 2009; Brooks et al. 2009; Ceverino, Dekel \& Bournaud 2010; Ceverino et al. 2014; Nuza et al. 2014a).

There has been some observational evidence for these accretion flows at high redshift, from the detection of inverse metallicity gradients (lower metallicity in the central star-forming regions than in the periphery of the sources; Cresci et al. 2010) and from high-column density \ion{H}{1} absorbers (e.g.,~van de Voort et al. 2012). However, in the local Universe, there has only been indirect evidence for these events, mainly from the study of star formation in dwarf galaxies. For example, for galaxies of the same mass, the metallicity was shown to decrease when star formation increases (Mannucci et al. 2010; Lara-L\'opez et al. 2010; P\'erez-Montero et al. 2013). It was also found that metallicities in quiescent blue compact dwarf (BCD) galaxies are higher than those of star-forming BCDs (S\'anchez Almeida 2008, 2009). More recently, observations of tadpole galaxies in the Kiso survey of UV-bright galaxies (Elmegreen et al. 2013) and extremely metal-poor dwarf star-forming galaxies (XMPs), have shown that the regions where star formation occurs possess a lower metallicity than the underlying galaxy (S\'anchez Almeida 2013, 2014a). From the \ion{H}{1} data, it has been shown that local BCDs and XMPs are generally surrounded by pools and streams of \ion{H}{1} gas, over three times larger than the optical extension, which frequently display, in the peripheries, distorted and lopsided \ion{H}{1} distributions and velocities (Begum \& Chengalur 2003; Young et al. 2003; Thuan, Hibbard \& L\'evrier 2004; Ekta, Chengalur \& Pustilnik 2006; Begum et al. 2006; Chengalur et al. 2006; Walter et al. 2008; Ekta, Chengalur \& Pustilnik 2008, 2009; Ekta \& Chengalur 2010; Ott et al. 2012; Lelli et al. 2012a, b; Lelli; Verheijan \& Fraternali 2014a, b; Beccari et al. 2014). Furthermore, significant quantities of metal-poor, large-scale \ion{H}{1} gas were found in the subset of XMPs (Filho et al 2013; hereinafter Paper~I). Because the gas in galaxy disks is diluted in a relatively small timescale (of the order of one galaxy rotation; e.g.,~de Avillez \& Mac Low 2002; Yang \& Krumholz 2012), this cumulative evidence suggests that the gas had to have been accreted recently, most likely as vestigial streams of cosmological accretion (S\'anchez Almeida 2014b), like the streams modeled in high-mass galaxies (Dekel et al 2009; Dekel, Sari \& Ceverino 2009; Ceverino, Dekel \& Bournaud 2010; Ceverino et al. 2014). The rationale is that the metallicity of the accreted \ion{H}{1} gas is low. As this gas is fed to the star formation regions of the source, the overall metallicity in this region is lowered, due to dilution of the gas, while elsewhere in the galaxy, the \ion{H}{1} gas has been enriched with metals from stellar winds and supernovae from previous star formation episodes.

Thus, XMPs may constitute the ideal local laboratories to probe the role of cosmological accretion flows in galaxy evolution and star formation. XMPs comprise over 0.1\% of the galaxies in the local volume (Morales-Luis et al. 2011), and represent the low-metallicity end of the most metal-poor galaxies, below an arbitrarily defined metallicity of one-tenth solar, or 12 + log(O/H) $\leq $ 7.65. Although most XMPs are a subsample of star-forming BCDs (Sargent \& Searle 1970; Thuan \& Martin 1981; Papaderos et al. 1996a; Papaderos et al. 1996b; Telles \& Terlevich 1997; Kunth \& \"Ostlin 2000; Cair\'os et al. 2001; Bergvall \& \"Ostlin 2002; Cair\'os et al. 2003; Noeske et al. 2003; Gil de Paz \& Madore 2005; Amor\'\i n et al. 2007, 2009; Micheva et al. 2013), some XMPs are more diffuse and correspond to the category of dwarf irregular galaxies (e.g., Skillman et al. 2013). They contain copious amounts of metal-poor, unsettled, \ion{H}{1} gas (Begum \& Chengalur 2003; Young et al. 2003; Thuan, Hibbard \& L\'evrier 2004; Ekta, Chengalur \& Pustilnik 2006; Begum et al. 2006; Chengalur et al. 2006; Walter et al. 2008; Ekta, Chengalur \& Pustilnik 2008, 2009; Ekta \& Chengalur 2010; Ott et al. 2012; Lelli et al. 2012a, b; Paper~I; Lelli, Verheijan \& Fraternali 2014a, b; Beccari et al. 2014). Moreover, in $\sim $ 80\% of the cases, XMPs show evidence for optical asymmetry (Papaderos et al. 2008; Morales-Luis et al. 2011; Paper~I). The lopsided optical morphology has been commonly interpreted as asymmetric star formation, occurring in a galaxy/disk that is in a transient phase (Papaderos et al. 2008; Elmegreen 2009; Elmegreen \& Elmegreen 2010), a rare occurrence at low redshift.

The objective of this paper (Paper~II) is to analyze the environment of the XMPs, with the aim of understanding the role of cosmological accretion flows in the galaxy/disk assembly and triggering/feeding of the star formation in XMPs. Archival high resolution interferometric \ion{H}{1} data are used to investigate the nearby environment, whereas cosmological simulations constrained by observations, and the observed distribution of galaxies, provide information on the environment on large scales. The sources are the same as in Paper~I, and haven been taken from the parent sample of 140 local XMPs, selected by Morales-Luis et al. (2011), from the Sloan Digital Sky Survey (SDSS) Data Release (DR) 7 (Abazajian et al. 2009) and literature. For 19 XMPs, archive \ion{H}{1} imaging, at resolutions higher than 1\arcmin, are available via interferometric Very Large Array (VLA) and/or Giant Metrewave Radio Telescope (GMRT) observations; these are used to probe the nearby environment. The cosmic web environment is inferred from the constrained numerical simulations by He{\ss}, Kitaura \& Gottl{\"o}ber (2013) and Nuza et al. (2014b), and the catalog of SDSS galaxy filaments assembled by Tempel et al. (2014). We have also searched for close companions to the XMPs in the SDSS database.

The paper is organized as follows. In Section~2 we present a summary of the optical and \ion{H}{1} properties of XMPs with archival high resolution \ion{H}{1} observations, which includes an analysis of the small-scale environment. Section~3 contains our investigation of the large-scale environment of XMPs in light of three diagnostic tools: cosmic web type, overdensity and nearest neighbours. In Section~4 we discuss the results and present the conclusions. The Appendix contains a qualitative description of the archival high resolution \ion{H}{1} morphology and velocity field of individual XMPs.

\section{Small-Scale Environment}

Of the 140 local XMPs in the parent sample, 29 sources with no \ion{H}{1} information have been observed with the single-dish Effelsberg radio telescope, resulting in 11 new detections (Paper~I). Overall, 53 XMPs possess \ion{H}{1} data in literature; but of these, only 19 have been observed at high spatial resolution ($\lesssim $ 1\arcmin), with either the GMRT or the VLA. We use these archival \ion{H}{1} observations to investigate the spatial and dynamical distribution of the \ion{H}{1} gas, and its relation to the optical component, which can provide clues as to the origin of the observed XMP properties.

\subsection{XMPs With Archival High Resolution \ion{H}{1} Observations}

Lelli, Verheijan \& Fraternali (2014a, b) have presented a detailed quantitative assessment of the \ion{H}{1} properties (from new and archival data) of 18 starburst dwarf galaxies, four of which are also in our XMP sample (UGC 4483, UGC 6456, IZw 18, and SBS 1415+437). In this paper, we attempt only a qualitative description of the high resolution \ion{H}{1} data in literature, in order to obtain global statistical properties for the XMP sample.

In Table~1 we present a summary of the optical properties of the sample of 19 local XMPs with archival high resolution interferometric \ion{H}{1} observations. The data is adapted from Table~3 of Paper~I, with some revision and description of the optical morphology. Table~2 provides a summary of the relevant \ion{H}{1} properties, at several scales, of the 19 local XMPs with archival high resolution interferometric \ion{H}{1} observations. The properties are a compilation of interferometric \ion{H}{1} data from literature, namely VLA and GMRT data. Low resolution maps display resolutions $\gtrsim $ 30\arcsec, while intermediate resolution display 30 -- 15\arcsec~and high resolution display $\lesssim $ 15\arcsec. In the Appendix, we present a detailed description of the \ion{H}{1} high resolution data for each XMP.

% PA using xfig on SDSS or DSS images.
% I define PA along the largest extension

%%%%%%%%%%%%%%%%%%%%%%%%%%%%%%%%%%%%%%%%%%%%%%%%%%%%%%%%%%%%%%%%%%%%

% TABLE 1 - Summary Optical Data

\setcounter{table}{0}

\begin{table*}

\footnotesize

\begin{center}

\begin{minipage}[c]{133mm}

\caption{Summary of the optical data for the 19 extremely metal-poor galaxies in the local Universe with archival high resolution interferometric HI observations. 
}

\begin{tabular}{l c c c c c c c c c c}

\hline

Source &   RA(J2000)        &   DEC(J2000)  	         & $m_{\rm g}$    &   $d_{\rm opt}$ & PA       & O/H    & Optical    & Note    \\
Name   &   $^h$ $^m$ $^s$   & $^\circ$  \, \arcmin \, \arcsec& mag    &   \arcsec       & $^\circ$ &        & Morphology &       \\
(1)  &  (2)               & (3)                          & (4)            &   (5)           &  (6)              & (7)   &   (8)   & (9)  \\

\hline
\hline
J0119-0935  &  01 19 14   & -09 35 46    		& 19.5       & 21.8 	&    115       & 7.31 	& cometary & \ldots \\
% classic SDSS comet
UM 133      &  01 44 42   & +04 53 42   	    & 15.4       & 54.7  	&    105       & 7.63   &  cometary & 1  \\	
% DSS comet curved at the end of the tail
SBS 0335-052W& 03 37 38   & -05 02 37    		& 19.0       & 5.0      & \ldots      & 7.11   &  symmetric & \ldots  \\	
% DSS blob no PA
SBS 0335-052E& 03 37 44   & -05 02 40    		& 16.3       & 8.3		& 45           & 7.31   &  cometary  & \ldots \\		
% DSS comet with faint NW tail
UGC 4305     &  08 19 05   & +70 43 12      	& \ldots    & 26.5		& 115 		   & 7.65   &  multi-knot disk & 2 \\
% DSS blob/disk but faint blobs to the NW and a loop-shape to the E, brightest knots on E edge of disk
HS 0822+03542& 08 25 55   & +35 32 31    		& 17.8       & 13.9     & 45           & 7.35    & cometary & \ldots \\	 	
% classic SDSS comet
DD 053      &  08 34 07   & +66 10 54           & 20.3		 & 71.2		 & 55	       & 7.62    &   multi-knot disk & 3  \\		
% SDSS multi-knot in a disk with a blob/disk to the W
UGC 4483    &  08 37 03   & +69 46 31    		& 15.1       & 61.8	     & 80		   & 7.58    &  cometary & 4 \\	
% DSS comet but X shape other fainter PA=60
IZw 18      &  09 34 02   & +55 14 25    		 & 16.4       & 18.5	 & 45          & 7.17    & two-knot cometary & 5 \\		
% SDSS comet with two-knots one on each end
Leo A       &  09 59 26   & +30 44 47           & 19.0       & 272.0     & 10          & 7.30     &  multi-knot disk & \ldots \\		
% SDSS multi-knot in a disk
Sextans B   &  10 00 00   & +05 19 56           & 20.5       & 269.0     & 25          & 7.53     & multi-knot disk & 6\\
% SDSS multi-knot in a disk with faint tail to W at PA=160
Sextans A   &  10 11 00   & -04 41 34           & \ldots    & 264.0     & 45  		   & 7.54     & multi-knot disk/symmetric & 7  \\	
% DSS blob/disk
UGC 6456      &11 28 00   & +78 59 39           & \ldots    & 93.0       & 115         & 7.35    & multi-knot disk & 8  \\
% DSS multi-knot in a disk but with a couple bright knots on the E edge 
SBS 1129+576  &11 32 02   & +57 22 46           & 16.7       & 45.6       & 75          & 7.36     &  cometary & \ldots \\
% classic SDSS comet
UGCA 292      &12 38 40   & +32 46 01           & 18.9      & 57.0	     & 165         & 7.28     & multi-knot disk  & 9 \\
% SDSS multi-knot in a boomerang-shape disk
GR 8           &12 58 40   & +14 13 03          & 17.9      & 78.0       & 130         & 7.65     & multi-knot disk &  10  \\	
% SDSS multi-knot in a disk but the disk shows a curve
SBS 1415+437  &14 17 01   & +43 30 05           & 17.8      & 55.5       & 120         & 7.43     & two-knot cometary & 11 \\ 
% SDSS comet with two-knots at the head and some curve
Sag Dig        &19 29 59   & -17 40 41   	    & \ldots   & 116.0      & 170         & 7.44     &  multi-knot disk/symmetric & 12  \\
% DSS blob/disk		
J2104-0035    &21 04 55   & -00 35 22           & 17.9     & 13.1       & 110         & 7.05     &  cometary & \ldots  \\
% classic SDSS comet
\hline

\end{tabular}
\\
Columns: 
\\
Col. 1: Source name. \\
Col. 2: Right Ascension (J2000). \\
Col. 3: Declination (J2000). \\
Col. 4: SDSS DR7 $g$-band (Petrosian) magnitude, from Morales-Luis et al. (2011). \\
Col. 5: Optical diameter along the major axis, obtained from the SDSS DR7 $r$-band or Digitized Sky Survey (DSS)-II $R$-band data, measured at 25 mag arcsec$^{-2}$. \\
Col. 6: Approximate optical position angle, along the largest extension, measured counterclockwise from the West, from the SDSS DR10 or DSS images.   \\
Col. 7: Metallicity, $12 \, + \, \log$ (O/H), from Morales-Luis et al. (2011). \\
Col. 8: Optical morphology, obtained from the SDSS DR10 or DSS images. Symmetric (overall spherically symmetric structure), disk (overall disk-like structure), two-knot (structure with two prominent star formation knots), multi-knot (structure with multiple star formation knots) and cometary (overall head-tail structure, with at least one prominent star formation knot), are used. \\
Col. 9: Note on the optical structure from the SDSS DR10 or DSS images.
\\
\\
Note in column 9:
\\
1 -- The extended cometary tail is curved to the West, in a "boomerang-type" structure. \\ 
2 -- There is a faint nodule of emission to the Northwest, a bright "loop-shape" structure to the East, and a faint filament, beginning near the loop, and curving to the Southwest. The brightest star formation knots are found on the Eastern edge of disk, near the loop and filament. \\
3 -- There is second disk-like emission to the West, at a position angle of approximately 55$^\circ$, similar to the main disk. The brightest star formation knots are found on the Northwestern tip of the main disk. \\
4 -- Although the overall shape is cometary, there appears to be a fainter X-shape pattern, with the brighter and the fainter arm at a position angle of approximately 95 and 60$^\circ$, respectively. \\
5 -- The two brightest star formation knots are each at the end of the cometary structure. \\
6 -- There is a faint tail of emission to the West, at a position angle of approximately 160$^\circ$. \\
7 -- The brightest star formation knots are found on the Eastern tip of the disk/symmetric structure.  \\
8 -- There is faint extended emission along a position angle of approximately 75$^\circ$. The brightest star formation knots are found on the Eastern edge of the disk. \\
9 -- The disk has a "boomerang" shape, with the brightest star formation knots in the center and on the Eastern tip of the disk. \\
10 -- The disk shows curvature at both ends, in a "boomerang-type" morphology. \\
11 -- The two brightest star formation knots are at the head of the cometary structure. The extended cometary tail is curved to the West. \\ 
12 -- The brightest star formation knots are found on the Eastern tip of the disk/symmetric structure.

\end{minipage}

\end{center}

\end{table*}

%%%%%%%%%%%%%%%%%%%%%%%%%%%%%%%%%%%%%%%%%%%%%%%%%%%%%%%%

% Position angles (optical, \ion{H}{1} and velocity gradient) are defined counterclockwise from the West, and are measured along the largest structural extension. \ion{H}{1}-to-optical ratios are measured using the lowest resolution \ion{H}{1} images. \ion{H}{1}-to-optical offsets are measured relative to the optical center and/or to the brightest star-forming knot. 

%%%%%%%%%%%%%%%%%%%%%%%%%%%%%%%%%%%%%%%%%%%%%%%%%%%%%%%%%%%

% TABLE 2 - Summary \ion{H}{1} Radio Data

\setcounter{table}{1}

\begin{table*}

\footnotesize

\begin{center}

\begin{minipage}[c]{126mm}

\caption{Summary of the \ion{H}{1} data of the 19 extremely metal-poor galaxies in the local Universe with archival high resolution interferometric HI observations.
}

\begin{tabular}{l c c c c c c r}

\hline

Source &   PA$_{\rm opt}$		&	HI/opt Ratio   & HI/opt Offset & PA$_{\rm HI}$    & 	PA$_{\rm VG}$	&	Note	&	Ref  	        \\
Name     &   $^\circ$	        &						   & \arcsec				& $^\circ$  			    & $^\circ$			&			&					\\
(1)  &  (2)                 & (3)                      & (4)          		    &   (5)          		&  (6)     	 		& (7)  		& (8)			\\

\hline
\hline

J0119-935		&	115		&	4		& 	10/10		&		120--110	 	&		120			&  \ldots		& 1 \\
UM 133			&	105		&	2		&	0/10		&		105				&		105			& \ldots		& 1 \\
SBS 0335-052W$^a$	& 	\ldots	&	6		& 	10/10		&		150				&		NE-SW--N-S	&	P			& 2 \\
SBS 0335-052E$^a$	&	45		&	4		&	10/10		& 		145--20			&		NE-SW		& 	P			& 2 \\
UGC 4305		&	115 	&	5		&	0/0			&		\ldots			&		100			& \ldots		& 3 \\	
HS 0822+03542	&	45		&	10		&	10/10		& 		70-80			& 		70			&   P			& 4 \\
DD 053			& 	55		&	3		&	10/10		&		150				&		45			& \ldots		& 3,5 \\
UGC 4483		&	80		&	2		&	10/0		&		80--60--80		&		90--120		& \ldots		& 6,7,8,9 \\
IZw 18			&	45		&	7		&	10/0		&		45				&		50--55		&   P			& 8,9,10 \\ 
Leo A			&	10		&	3		&	\ldots		&		10				&		\ldots		& \ldots		& 5 \\
Sextans B		&	25		&	3		&	10/10		&		25				&		130			& \ldots		& 6 \\
Sextans A		&	45		&	2		&	10/0		&		135				&		150			& \ldots		& 6 \\
UGC 6456		&	115		&	2		&	10/0		&		75				&		90			& \ldots		& 8,9,11 \\
SBS 1129+576	&	75		&	2		&	0/0			&		75--70			&		70			&	P			& 12 \\
UGCA 292		&	165		&	3		&	0/0			&		175				&		145			& \ldots		& 6, 13 \\
GR 8			&	130		&	2		&	10/10		&		165				&		80			& \ldots		& 5,6,13,14 \\
SBS 1415+437	&	120		&	2		&	10/0		&		115				&		115			& \ldots		& 8,9 \\
Sag Dig			&	170		&	6		&	0/0			&		\ldots			&		\ldots		&   M 			& 5, 15 \\
J2104-0035		&	110		&	5		&	10/10		&		90				&		90			& \ldots		& 16 \\
\hline

\end{tabular}
\\
$^a$SBS 0335-052E and SBS 0335-052W are an interacting pair. Due to lack of resolution, the PA$_{\rm VG}$ is merely indicative. 
\\
\\
Columns: \\
Col. 1: Source name. \\
Col. 2: Approximate position angle of the largest optical extension, measured counterclockwise from the West, from the SDSS DR10 or DSS images. \\
Col. 3: Approximate HI-to-optical ratio, where the HI extension is measured in the lowest resolution interferometric map. \\
Col. 4: Approximate offset between the HI peak and the optical center/brightest star-forming knot. \\
Col. 5: Approximate position angle of the largest HI extension, measured counterclockwise from the West (from the lowest to highest resolution map). \\
Col. 6: Approximate position angle of the HI velocity gradient, measured counterclockwise from the West (from the lowest to highest resolution map). \\
Col. 7: Note on the possible interaction/merger of the XMP. "P" designates a galaxy that belongs to a confirmed interacting pair. "M" designates a galaxy that may have undergone a recent merger. \\
Col. 8: Reference, with interferometer data and/or survey.
\\
\\
Instrument, interferometer and survey acronym:
\\
GMRT -- Giant Metrewave Radio Telescope \\
VLA -- Very Large Array \\
THINGS -- The HI Nearby Galaxy Survey \\
ACS -- Advanced Camera for Surveys \\
ANGST -- ACS Nearby Galaxy Survey Treasury \\
\\
Reference in column 8: 
\\
1 -- Ekta \& Chengalur (2010); GMRT \\
2 -- Ekta, Pustilnik \& Chengalur (2009); GMRT \\
3 -- Walter et al. (2008); VLA THINGS \\
4 -- Chengalur et al. (2006); GMRT \\
5 -- Begum et al. (2006); GMRT \\
6 -- Ott et al. (2012); VLA ANGST \\
7 -- Lelli et al. (2012b); VLA  \\
8 -- Lelli, Verheijan \& Fraternali (2014a); VLA \\
9 -- Lelli; Verheijan \& Fraternali (2014b); VLA \\
10 -- Lelli et al. (2012a); VLA \\
11 -- Thuan, Hibbard \& L\'evrier (2004); VLA \\
12 -- Ekta, Chengalur \& Pustilnik (2006); GMRT \\
13 -- Young et al. (2003); VLA \\
14 -- Begum \& Chengalur (2003); VLA \\
15 -- Beccari et al. (2014); VLA \\
16 -- Ekta, Chengalur \& Pustilnik (2008); GMRT

\end{minipage}

\end{center}

\end{table*}

%%%%%%%%%%%%%%%%%%%%%%%%%%%%%%%%%%%%%%%%%%%%%%%%%%%%%%%%%%%%%%%%%

%We have adopted the classification of Lelli et al. (2014a) to characterize the \ion{H}{1} distribution and kinematics. Type A are sources with velocity fields typical of a rotating disk, which may present small asymmetries, and whose \ion{H}{1} peak may or may not coincide with the optical center. Type B are sources which show asymmetries and irregularities in the velocity fields, although the \ion{H}{1} distribution resembles a disk. Type C are sources which show asymmetries and distortions both in the \ion{H}{1} distribution and kinematics.} We find that most of the XMPs are Type A sources.

% The outer contours \ion{H}{1} show clear irregularities, with tails, extensions, bridges and ridges.

If we compare the XMP \ion{H}{1} morphology and velocity field with the classification given in Lelli, Verheijan \& Fraternali (2014b), we conclude that almost all of the XMPs fall within their classification of Type A/B sources -- galaxies with regularly rotating \ion{H}{1} disks/galaxies with kinematically disturbed \ion{H}{1} disks. That is, the XMPs generally show an \ion{H}{1} disk-type morphology, with a velocity field typical of a rotating disk, which may present some small asymmetries and irregularities in the outskirts, and which may be (only slightly) offcenter relative to the optical peak and/or star-forming knots; statistically, we do not observe significant perturbations in the large-scale \ion{H}{1} gas component of the XMP population (Appendix). High spatial resolution observations of the \ion{H}{1} gas suggest feeding of the star formation regions; the gas is resolved into complex substructure, with the gas "pointing at" or "cradling" bright star-forming regions, or showing signs of "clearing" around the star-formation knots (Appendix). Sag Dig has been suggested to be the result of a merger process (Beccari et al. 2014 and Appendix). Its \ion{H}{1} morphology is ring-like, with an offcenter depression, and the velocity field displays no sign of rotation. These features are generally not found in XMPs, which we interpret as evidence that the XMPs, as a class, are currently not undergoing a major interaction or merger. Rather, the asymmetries and irregularities of the outer envelope are understood as signs of weaker interactions caused by minor mergers, gas accretion or gas outflows driven by galactic winds. 

\section{Large-Scale Environment}

%In numerical simulations, the environment of a model galaxy or group of galaxies can be probed using a gravitational tidal-tensor-based structure-finding algorithm (Hahn et al. 2007b; Forero-Romero et al. 2009) to characterize the large-scale cosmic web region in which it resides\footnote{Voids, sheets and filaments correspond to expansions in three, two and one dimension, respectively. Knots represent pure contraction.} (voids, sheets, filaments and knots). In observations, the environment can be investigated by estimating the overdensity, defined as the number density of galaxies relative to the background number density, or by the characterization of the nearest neighbours (type, mass, number and distance).

In order to investigate the environment in which XMPs occur, and the influence that cosmological accretion flows may have on their evolution and observed properties, we have computed the likelihood of an XMP to inhabit a certain type of environment using three diagnostics: cosmic web type, overdensity and nearest neighbours. 

The environment can be classified according to local indicators, such as the local density, or according to non-local indicators, based, for example, on the tidal-field tensor. In the first case, one can estimate the local overdensity based solely on observations, defined as the number density of galaxies relative to the background number density, or by characterizing the nearest neighbours (type, mass, number and distance). However, a classification of the environment into its cosmic web components requires knowledge of the large-scale environment. The sign of the eigenvalues of the tidal-field tensor indicate pure contracting regions, such as clusters or knots, or expanding regions in three, two, or one dimensions, representing voids, sheets, or filaments, respectively (Hahn et al. 2007). While numerical simulations provide all the necessary information to perform a cosmic web analysis, observations require a prior reconstruction of the large-scale matter density field. The particular reconstruction technique will determine the accuracy of the cosmic web classification. 

%Let us describe our approach below.

%The T-web method (Hahn et al. 2007b; Forero-Romero et al. 2009) makes use of the gravitational tidal tensor structure-finding algorithm on a series of high resolution constrained N-body simulations to investigate large-scale structure in the Universe. 

\subsection{Cosmic Web Type and Overdensity}

%The cosmological model provides a statistical description of the Universe, consistent with an infinite number of realizations. Nuza et al. (2014b) generate a large series of cosmological N-body numerical simulations, compatible with the concordance Universe, chosing the one closest to the observations, as traced by the galaxies in the zero redshift Two-Micron All-Sky Galaxy Redshift Survey (2MRS; Huchra et al. 2012). This approach allows the authors to reconstruct a structure of volume 180$^{3}h^{-3}$ Mpc$^3$, that resembles the observed local Universe, with a spatial resolution of $\sim $ 2~Mpc. The overdensity ($\delta $) of this model of the local Universe has been computed, as determined directly from the reconstruction. By measuring the tidal-field tensor (Hahn et al. 2007b; Forero-Romero et al. 2009) from the reconstructed mass distribution, it is possible to perform a cosmic web characterization by type (see Nuza et al. 2014b for details). 

We here rely on precise constrained N-body simulations of the Local Universe, using the Two-Micron All-Sky Galaxy Redshift Survey (2MRS; Huchra et al. 2012), within volumes of 180~$h^{-3}$~Mpc$^3$ (He{\ss}, Kitaura \& Gottl{\"o}ber 2013). Cosmological parameters only provide a statistical description of the Universe, so that an infinite number of realizations are consistent with the same set of parameters. The procedure provides a cosmological simulation, which reproduces, in detail, the spatial distribution of galaxies in the local universe as described by 2MRS. The method to reconstruct the initial fluctuations used in these simulations is based on a self-consistent Bayesian machine-learning algorithm, reaching an accuracy of $\sim$ 2~Mpc~$h^{-1}$ (Kitaura et al. 2012; Kitaura 2013; He{\ss}, Kitaura \& Gottl{\"o}ber 2013). The cosmic web classification used in this study has been performed following Forero-Romero et al. (2009), based on the best correlated constrained N-body simulation with the 2MRS galaxy field (Nuza et al. 2014b). The overdensity ($\delta$) used in this study has been computed from the same constrained simulation. The preference of a type of galaxy to be located in a particular environment is quantified by $\eta$ ($\tau $, $\epsilon$), an excess probability ratio, where $\tau $ is the galaxy type and $\epsilon $ is the cosmic web environment type (voids, sheets, filaments and knots; Nuza et al. 2014b). The environment (overdensity or cosmic web type) of a source can be assigned according to the cell in the reconstruction where the source is located. Each cell is characterized by three coordinates: right ascension, declination, and redshift, with the redshift parameterizing the distance, including the proper motions caused by local tidal fields, taken consistently into account.

Because we possess coordinates for our XMP galaxies, we have used the constrained simulation to determine the type of environment (comic web type and overdensity) in which the XMPs reside. Spectroscopic redshifts from the SDSS DR10 (Ahn et al. 2012) are available for $\sim $ 70\% of the 140 XMPs. For the remaining XMPs, photometric redshifts from the SDSS DR10 or NASA/IPAC Extragalactic Database\footnote{http://ned.ipac.caltech.edu/} (NED) were used. 

In Figure~1, we have compared the cosmic web type environment distribution of XMPs with other galaxy morphological types, namely, ellipticals (E), lenticulares (S0), spirals (Sp) and irregular galaxies (Irrs), as assigned by the 2MRS team (Huchra et al. 2012). Although the Poisson errorbars are larger for the XMP sample, it is apparent that the XMPs follow similar trends as Irr galaxies, but with a more pronounced behaviour (Fig.~1). The XMP prevalence in voids and sheets, and the avoidance of knots, suggests that, statistically and on a large scale ($\sim $ 1~Mpc), the XMPs are relatively isolated (Fig.~1).

%%%%%%%%%%%%%%%%%%%%%%%%%

% Figure 1 - Comparison XMPs with other galaxies in 2MRS

\setcounter{figure}{0}
\begin{figure*}
\begin{center}
\includegraphics[width=6cm]{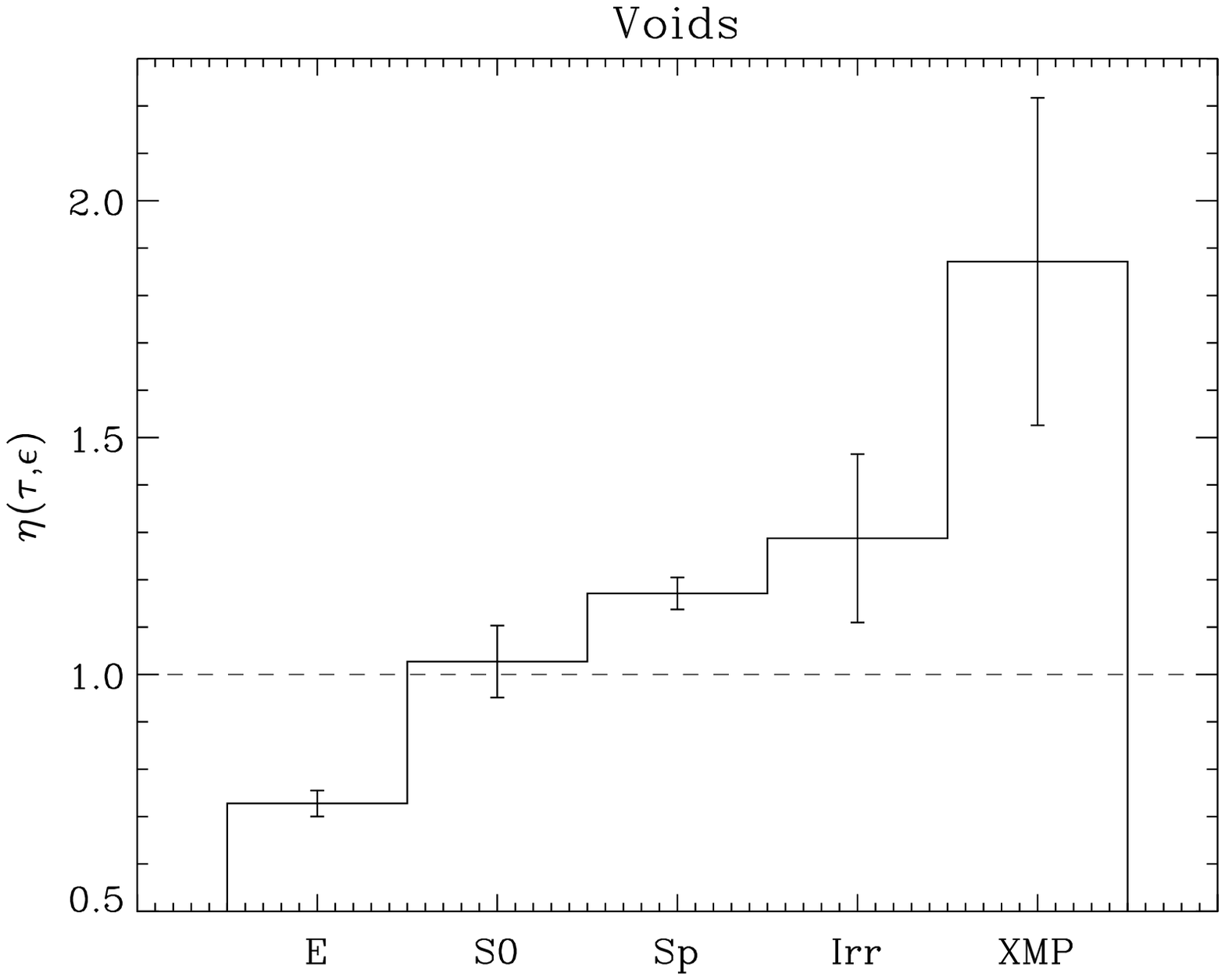}
\includegraphics[width=6cm]{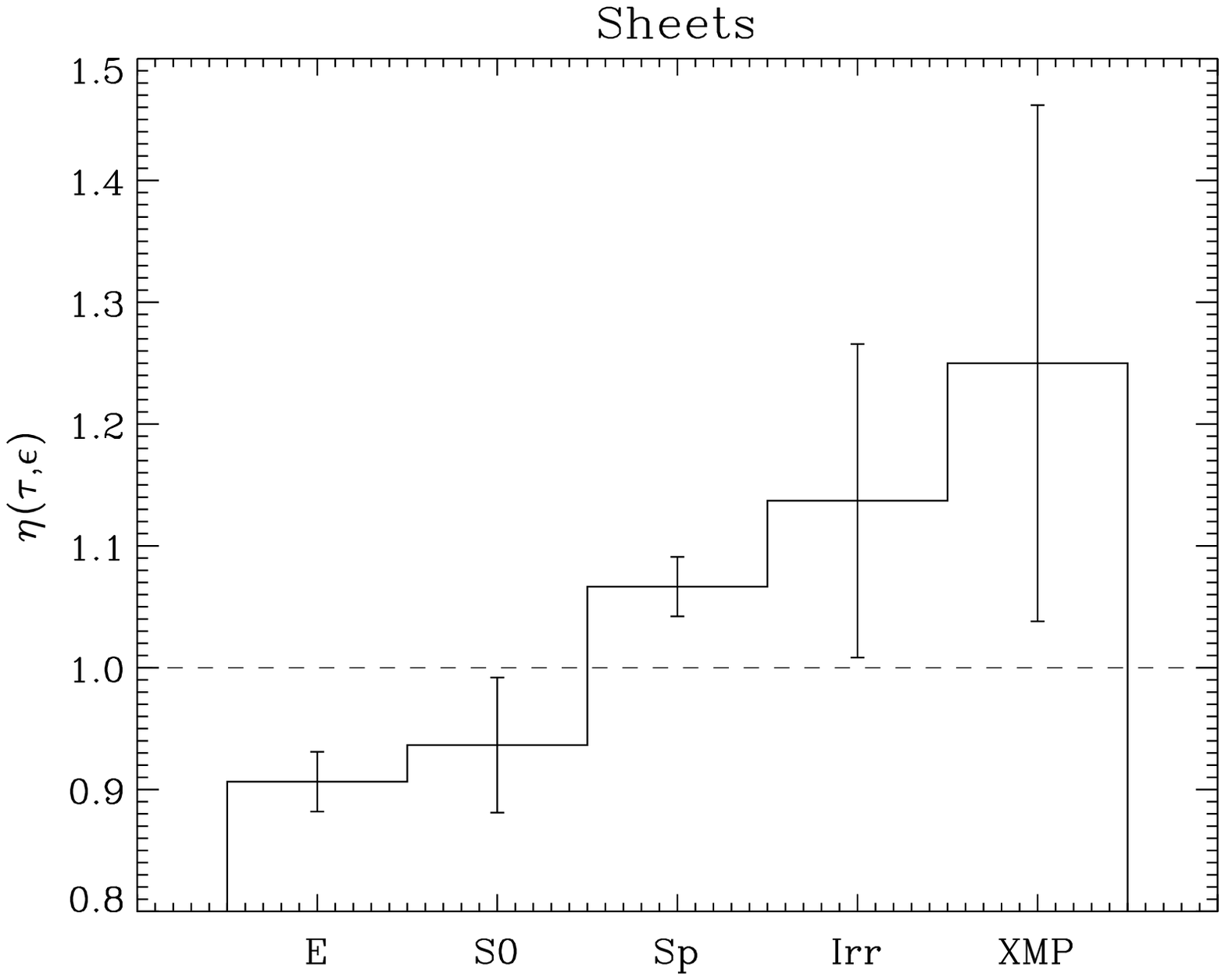}\\
\includegraphics[width=6cm]{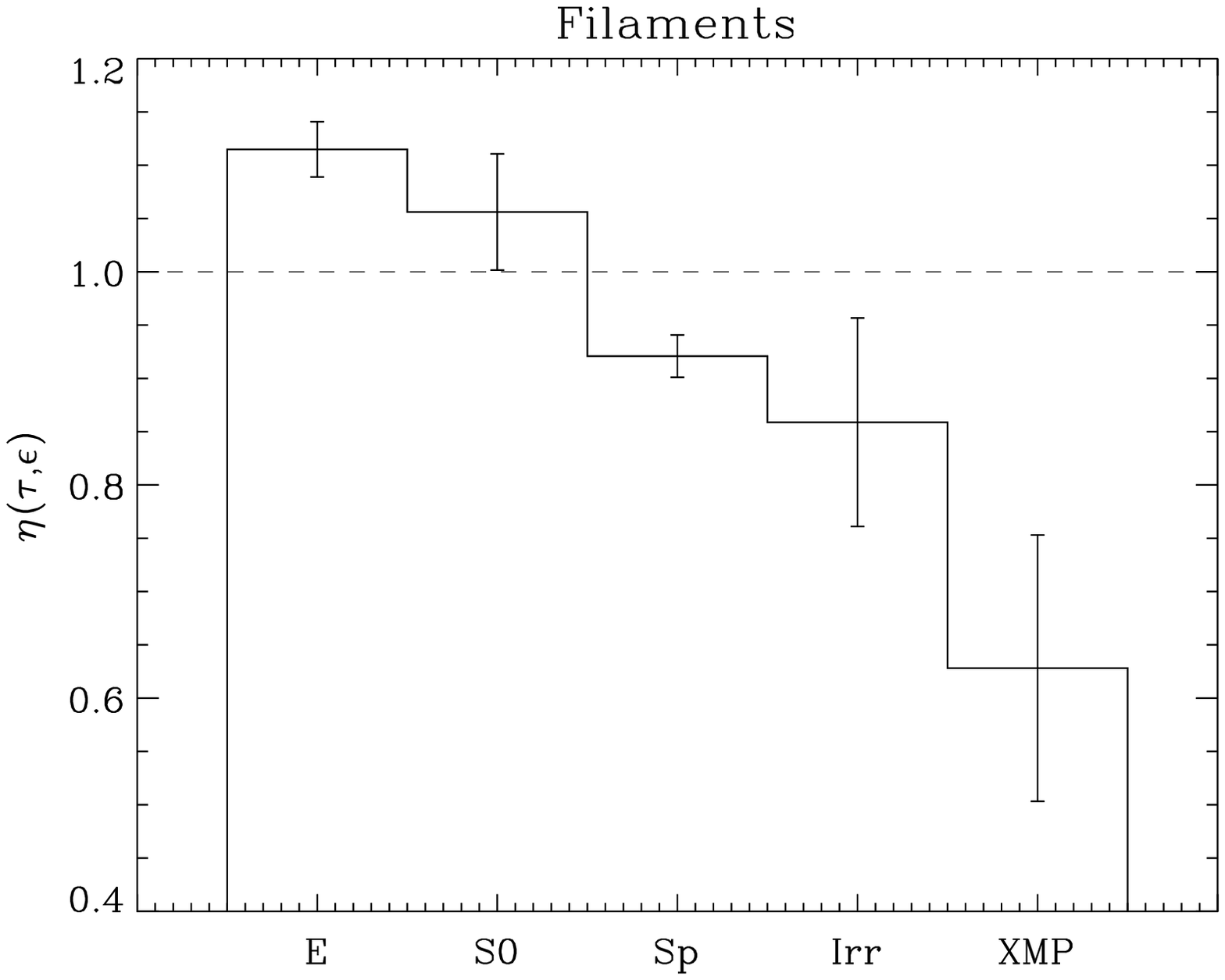}
\includegraphics[width=6cm]{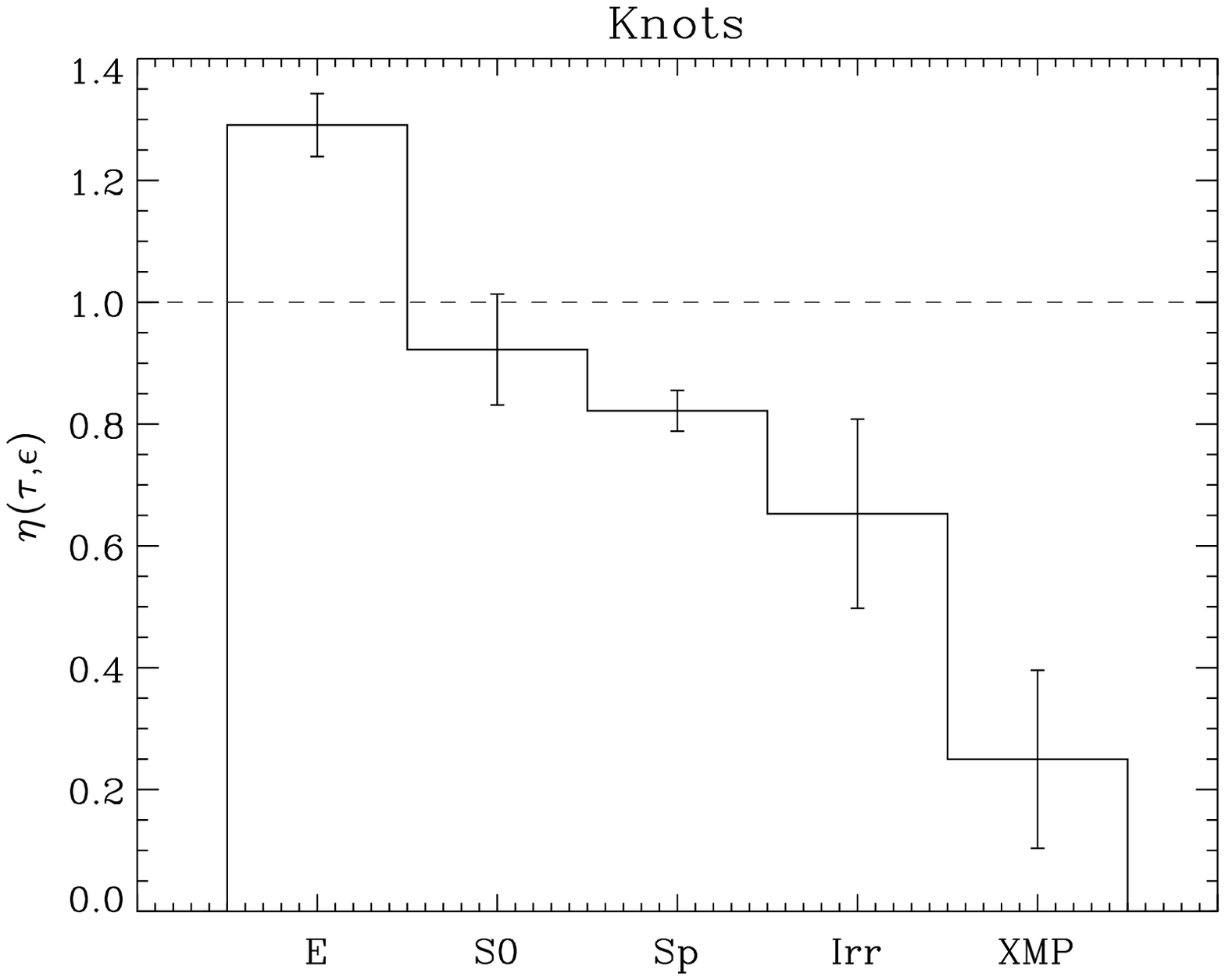}
\caption{The excess probability distribution of XMPs, and E, S0, Sp and Irr galaxies of the zero redshift 2MRS sample (Huchra et al. 2012; Nuza et al. 2014b). The four panels correspond to sources found in voids, sheets, filaments and knots.}
\end{center} 
\end{figure*}

%%%%%%%%%%%%%%%%%%%%%%%%%%%

As a consistency check, we have also estimated the cosmic web type distribution for a large sample of control galaxies ($\sim $ 700 BCDs, with Z  $>$ 0.1 Z$_{\odot}$) from the SDSS DR8 (Aihara et al. 2011). We find a similar behaviour among the BCDs and XMP sample, which is markedly different from the environments found for E, S0 and Sp galaxies, and more extreme than the behaviour of the Irr galaxies in the zero redshift 2MRS sample (to be included in a future paper).  

%%%%%%%%%%%%%%%%%%%%%%%%

% Figure 1 - XMP positions, projected on the reconstructed cosmic web

%\setcounter{figure}{0}

%\begin{figure*}
%\begin{center}
%\includegraphics[width=6cm]{/home/mfilho/WORK/NEW_XMPs/Environment/Plots_Sebas_NEW/voids_rec.ps}
%(a)
%\includegraphics[width=6cm]{/home/mfilho/WORK/NEW_XMPs/Environment/Plots_Sebas_NEW/sheets_rec.ps}
%(b)
%\includegraphics[width=6cm]{/home/mfilho/WORK/NEW_XMPs/Environment/Plots_Sebas_NEW/filaments_rec.ps}
%(c)
%\includegraphics[width=6cm]{/home/mfilho/WORK/NEW_XMPs/Environment/Plots_Sebas_NEW/knots_rec.ps}
%(d)
%\includegraphics[width=6cm]{/home/mfilho/WORK/NEW_XMPs/Environment/Plots_Sebas_NEW/dens_rec.ps}
%(e)
%\caption{The projected positions, in supergalactic coordinates, onto a 7$h^{-1}$~Mpc slice of the reconstruction, of $\sim$ 45 XMPs. The different web types (a--d), as well as the dark matter overdensity field (e), are represented by white regions. XMPs (filled circles) are color-coded by cosmic web type (blue for voids, red for sheets, green for filaments and cyan for knots).}
%\end{center} 
%\end{figure*} 

%%%%%%%%%%%%%%%%%%%%%%%%

Figure 2 shows the projected spatial distribution of the 140 local XMPs, color-coded for cosmic web type. The sizes of the circles trace the overdensity ($\delta $), such that, the larger the circle, the larger the overdensity. The values range from the most underdense ($\delta $ = $-$0.9, for PHL 293B) to the most overdense ($\delta $  = 376, for VCC 0428, a BCD in the Virgo cluster; Meyer et al. 2014) source environment. Figure 3 shows the overdensity distribution (log(1+$\delta $)), color-coded by cosmic web type, for the XMP sample. Arrows, color-coded by cosmic web type, show the mean overdensity (log(1 + $<\delta >$)) for each distribution. According to the overdensity distribution, $\sim $ 60\% of the XMPs are found in underdense regions ($\delta \lesssim $ 0) of the reconstruction (Fig.~2 and 3). On the other hand, most of the XMP galaxies ($\sim $ 75\%) reside in the less dense cosmic web type environments ($\sim $ 40\% are in sheets and $\sim $ 34\% in voids), while $\sim $ 25\% can be found in filaments, and $\lesssim $ 1\% in knots (Fig.~1, 2 and 3). 

%%%%%%%%%%%%%%%%%%%%%%%%

% Figure 2 - XMP positions in the sky, color-coded by cosmic web type

\setcounter{figure}{1}

\begin{figure} 
\begin{center}
\includegraphics[width=6cm]{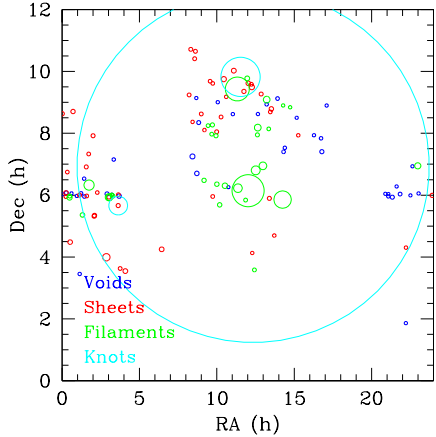}
\caption{The projected spatial distribution of the 140 local XMPs, color-coded by cosmic web type (blue for voids, red for sheets, green for filaments and cyan for knots). The sizes of the circles trace the overdensity, such that, the larger the circle, the larger the overdensity. The values range from the most underdense ($\delta $ = $-$0.9, for PHL 293B) to the most overdense ($\delta $ = 376, for VCC 0428, a BCD in the Virgo cluster) source environment.}
\end{center}
\end{figure} 

%%%%%%%%%%%%%%%%%%%%%%%%%%%

%%%%%%%%%%%%%%%%%%%%%%%%

% Figure 3 - XMPs and 2MRS overdensity histogram, color- coded by cosmic web type

\setcounter{figure}{2}

\begin{figure}
\begin{center}
\includegraphics[width=6cm]{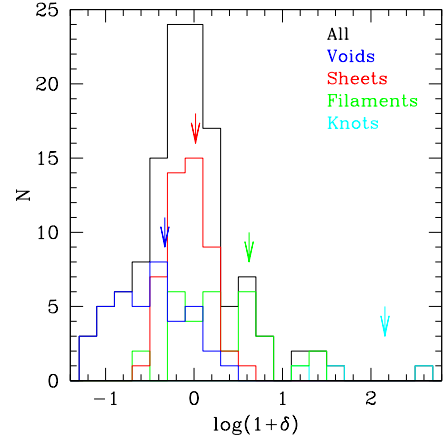}
\caption{The overdensity distribution for the 140 local XMP sample, color-coded by cosmic web type (blue for voids, red for sheets, green for filaments and cyan for knots). Arrows, color-coded by cosmic web type, show the mean overdensity (log(1 + $<\delta >$)) for each distribution.}
\end{center}
\end{figure}

%%%%%%%%%%%%%%%%%%%%%%%%%%%

%With respect to the relation between the overdensity and the cosmic web type (Fig.~3a), although there is a considerable overlap, we observe a global trend, whereby, as we move from the sequence knots $\rightarrow $ filaments $\rightarrow$ sheets $\rightarrow$ voids, we probe less dense environments. Indeed, the XMPs show mean overdensities ($<\delta >$) of $-$0.53 for voids, 0.04 for sheets, 3.19 for filaments and 143.16 for knots (Fig.~3a). {\bf This behaviour is also observed in the zero redshift 2MRS sample, which show mean overdensities of $-$0.51 for voids, $-$0.03 for sheets, 4.10 for filaments and 375.65 for knots (Fig.~3b).}

% For 2MRS
%Mean delta for galaxies in knots     :      375.653
%Mean delta for galaxies in filaments :      4.09965
%Mean delta for galaxies in sheets    :   -0.0323263
%Mean delta for galaxies in voids     :    -0.511029

% For XMPs
%Mean delta for galaxies in knots     :      143.156662
%Mean delta for galaxies in filaments :      3.190414667s
%Mean delta for galaxies in sheets    :   0.03950652853
%Mean delta for galaxies in voids     :    -0.5324053764

%%%%%%%%%%%%%%%%%%%%%%%%%%%

In the analysis below, because we are constrained by small number statistics and systematic errors that may affect the determination of the physical parameters, the trends should be taken with caution. 

We have investigated whether the \ion{H}{1} mass (M$_{\rm HI}$), dynamical mass (M$_{\rm Dyn}$) or stellar mass (M$_{\ast}$) of the XMPs depends on the environment in which they reside (Fig.~4). The \ion{H}{1} and stellar masses were taken from Table~5 of Paper~I. In this Paper~II, we have estimated the dynamical masses with revised optical radius measurements. There is a large overlap in dynamical, \ion{H}{1} and stellar mass for a range of overdensities (Fig.~4). However, XMPs in filaments and sheets show the largest range in \ion{H}{1} mass, particularly extending to lower \ion{H}{1} masses, while XMPs in voids show tendentiously larger \ion{H}{1} masses, above log (M$_{\rm HI}$/M$_{\odot}$) $\simeq $ 8 (Fig.~4a). Five sources in voids and sheets show the largest dynamical masses (log (M$_{\rm Dyn}$/M$_{\odot}$) $\gtrsim $ 9; Fig.~4b). In terms of stellar mass, most XMPs in filaments found in overdense regions ($\delta \gtrsim $ 0) show stellar masses below log (M$_{\ast}$/M$_{\odot}$) $\simeq $ 7, with a tail of XMPs in less dense filamentary environments towards high stellar masses (log (M$_{\ast}$/M$_{\odot}$) $\gtrsim $ 8; Fig.~4c). The largest \ion{H}{1}-to-stellar mass ratio belongs to an XMP residing in a void (Fig.~4d). Overall, we see no clear relation between the environment in which XMPs reside and their dynamical, stellar or \ion{H}{1} mass.

%%%%%%%%%%%%%%%%%%%%%%%%%%%%%

% Figure 4 - XMPs overdensity/cosmic web type as a function of mass

\setcounter{figure}{3}

\begin{figure*}
\begin{center}
\includegraphics[width=6cm]{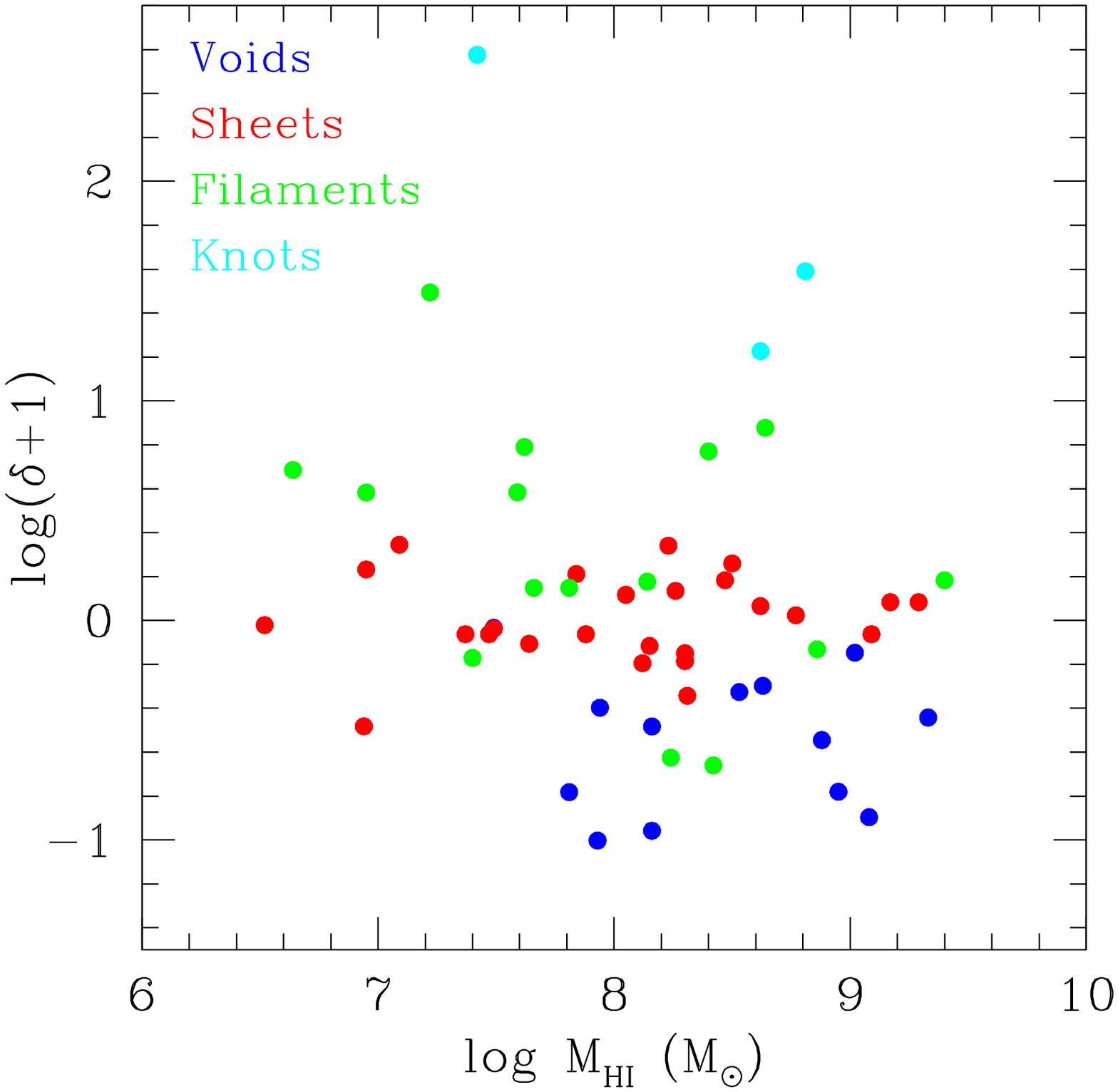}
(a)
\includegraphics[width=6cm]{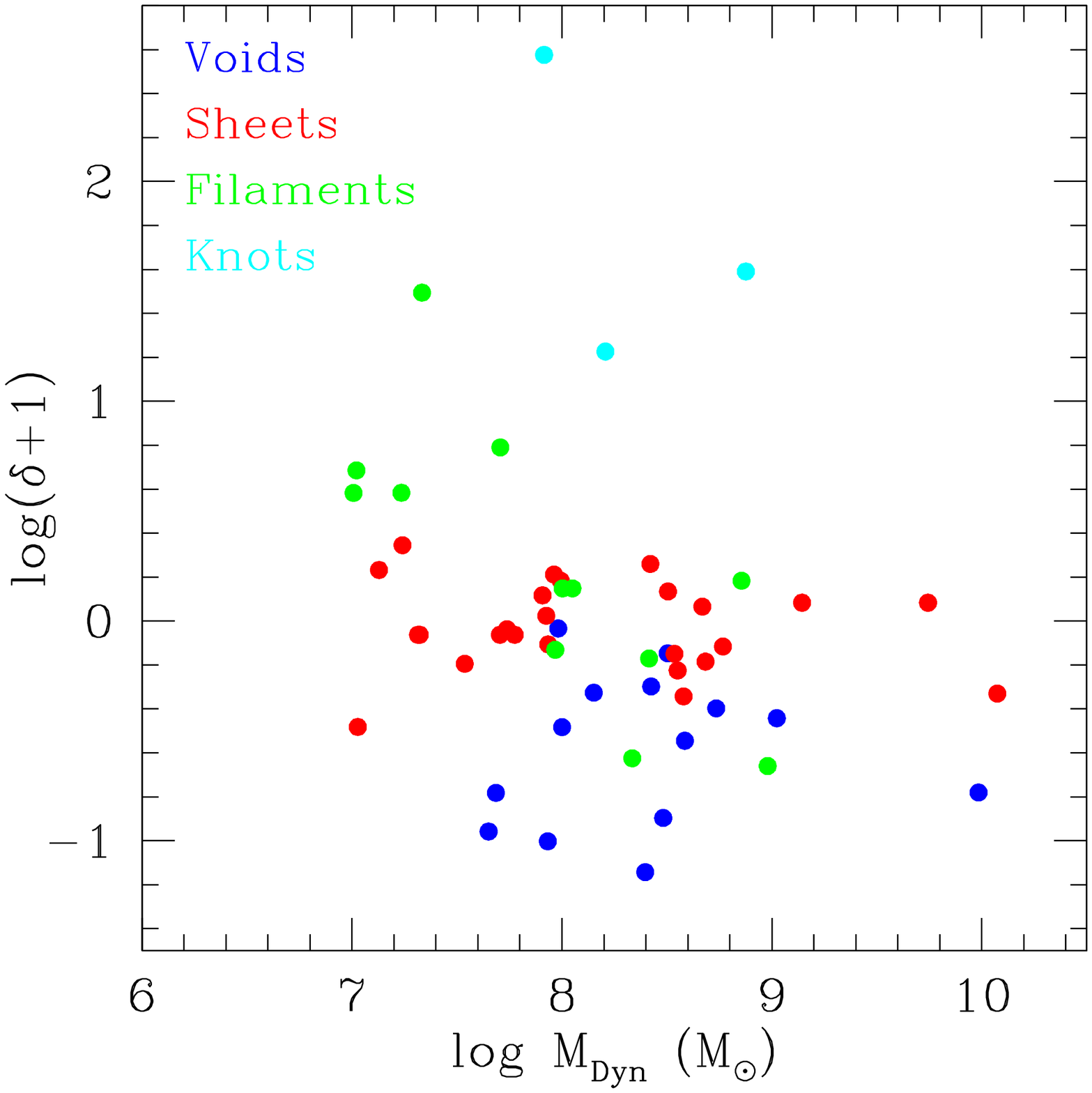}
(b) \\
\includegraphics[width=6cm]{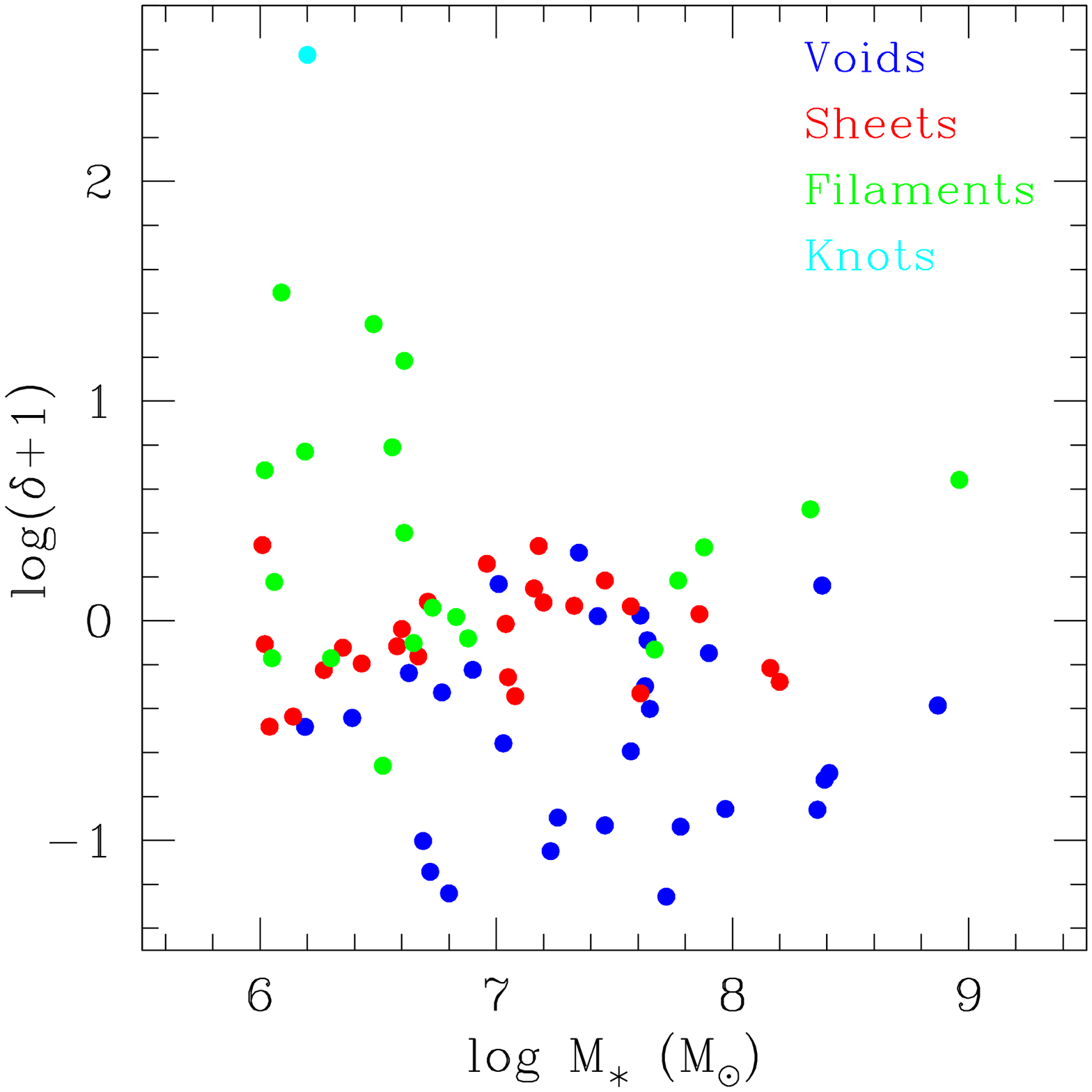}
(c)
\includegraphics[width=6cm]{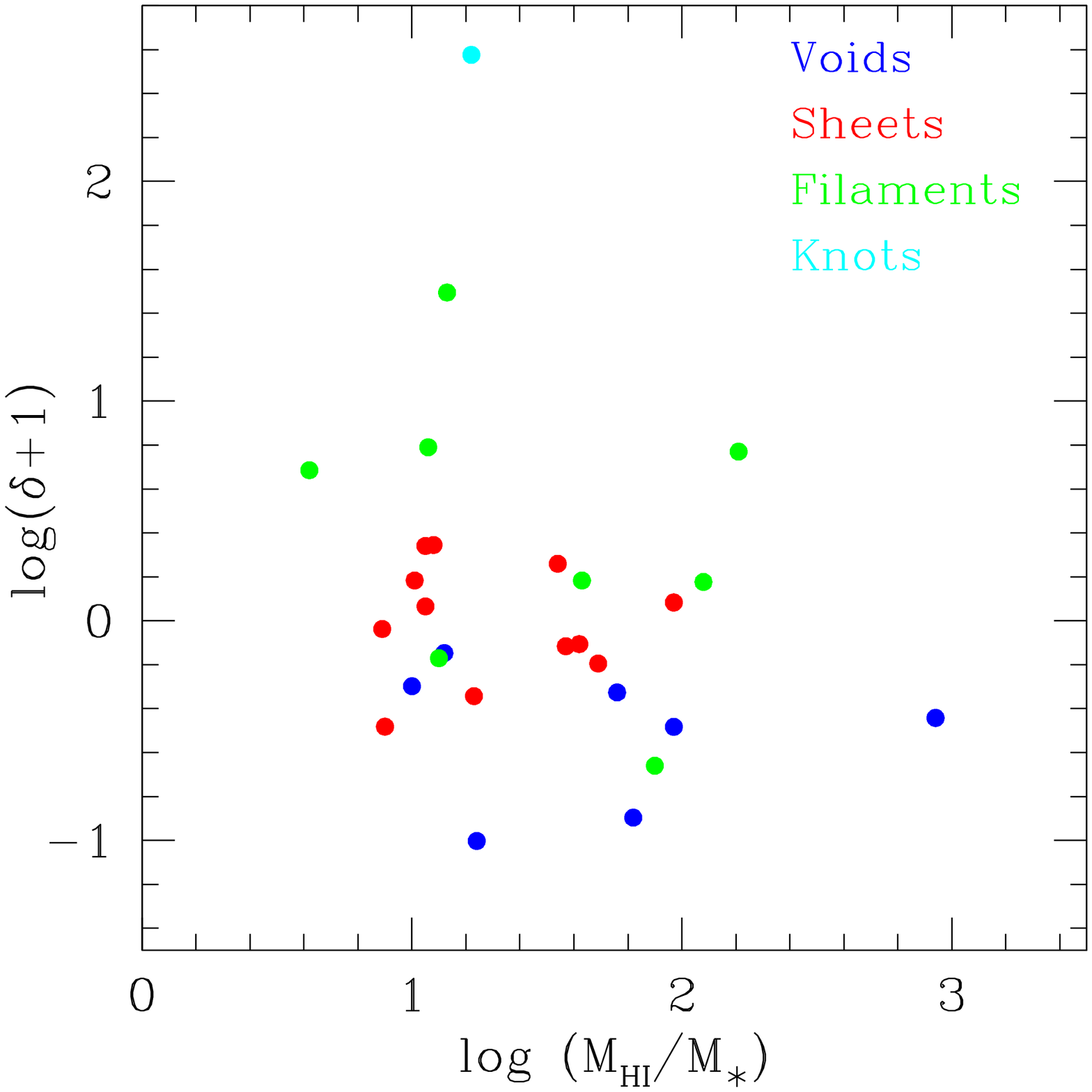}
(d)
\caption{The overdensity as a function of (a) \ion{H}{1} mass, (b) dynamical mass, (c) stellar mass, and (d) \ion{H}{1}-to-stellar mass ratio, color-coded by cosmic web type (blue for voids, red for sheets, green for filaments and cyan for knots). The \ion{H}{1} and stellar masses were taken from Table~5 of Paper~I. In this Paper~II, we have estimated the dynamical masses with revised optical radius measurements.}
\end{center} 
\end{figure*} 

%%%%%%%%%%%%%%%%%%%%%%%%%%%

We have further investigated if there is an environment trend with metallicity (12 + log(O/H)), star formation rate (SFR), specific star formation rate (sSFR $\equiv $ SFR/M$_{\ast}$), \ion{H}{1} gas consumption rate (1/t$_{\rm M_{\rm HI}} \equiv $ SFR/M$_{\rm HI}$), and optical morphology (symmetric, cometary, two-knot and multi-knot; Fig.~5). These values were taken from Table~3 and 4 of Paper~I. We find no clear correlation between environment and metallicity (Fig.~5a), SFR (Fig.~5b) and gas consumption rate (Fig.~5d). However, XMPs in voids show a slight tendency for larger values of sSFR (log sSFR $\gtrsim $ $-$8 yr$^{-1}$; Fig.~5c). Cometary and multi-knot XMPs are roughly equally (fractionally) prevalent in voids, sheets and filaments (Fig.~5e and 5f), while two-knot sources are roughly equally (fractionally) prevalent in voids and sheets (Fig.~5e and 5f). Symmetric sources are found to be more common in voids (Fig.~5e and 5f). The two sources with morphological information, and contained in knots, are both cometary (Fig.~5d and 5e).

%%%%%%%%%%%%%%%%%%%%%%%%

% Figure 5 - XMPs overdensity/cosmic web type as a function of other properties

\setcounter{figure}{4}

\begin{figure*}
\begin{center}
\includegraphics[width=6cm]{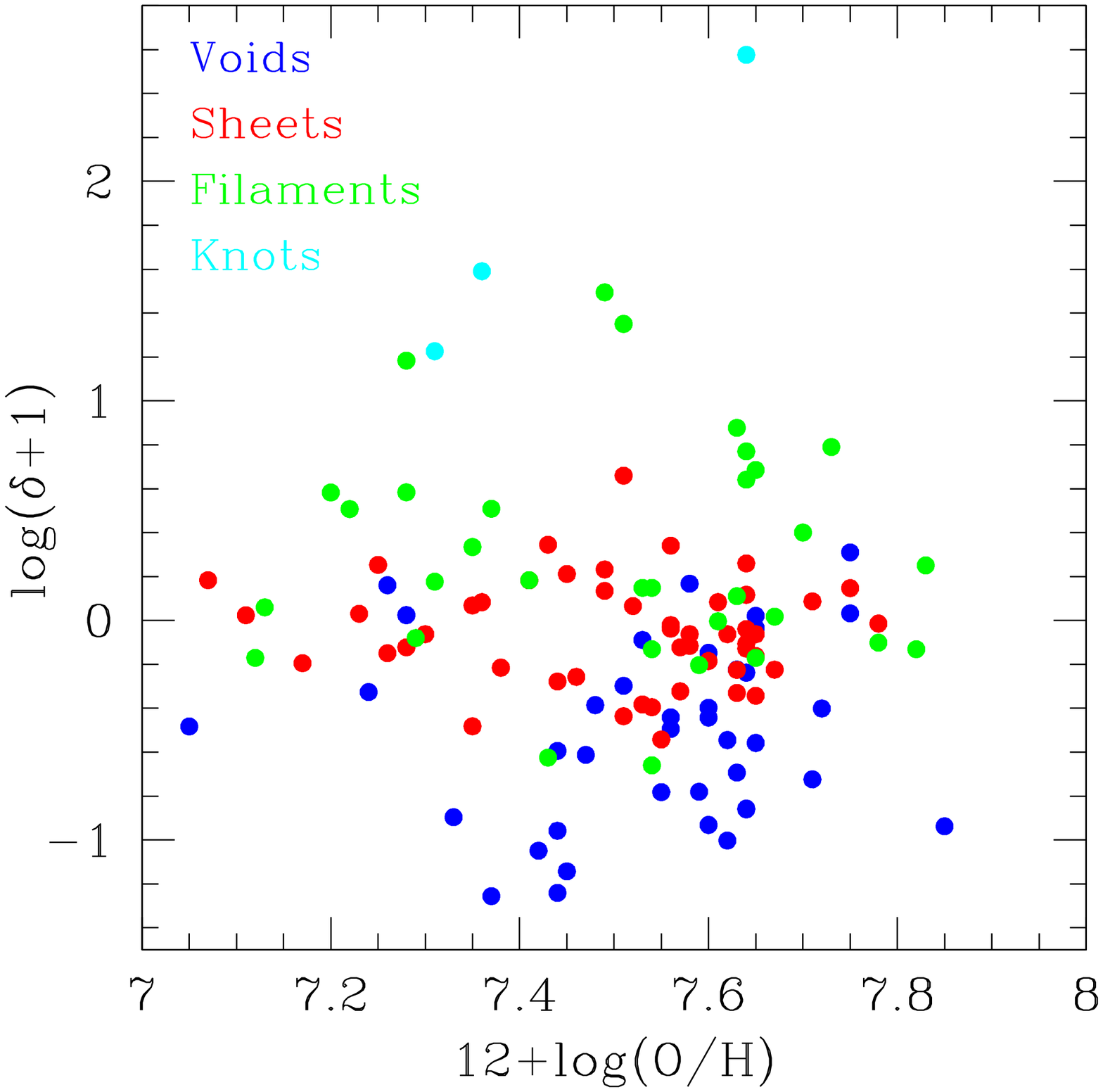}
(a) 
\includegraphics[width=6cm]{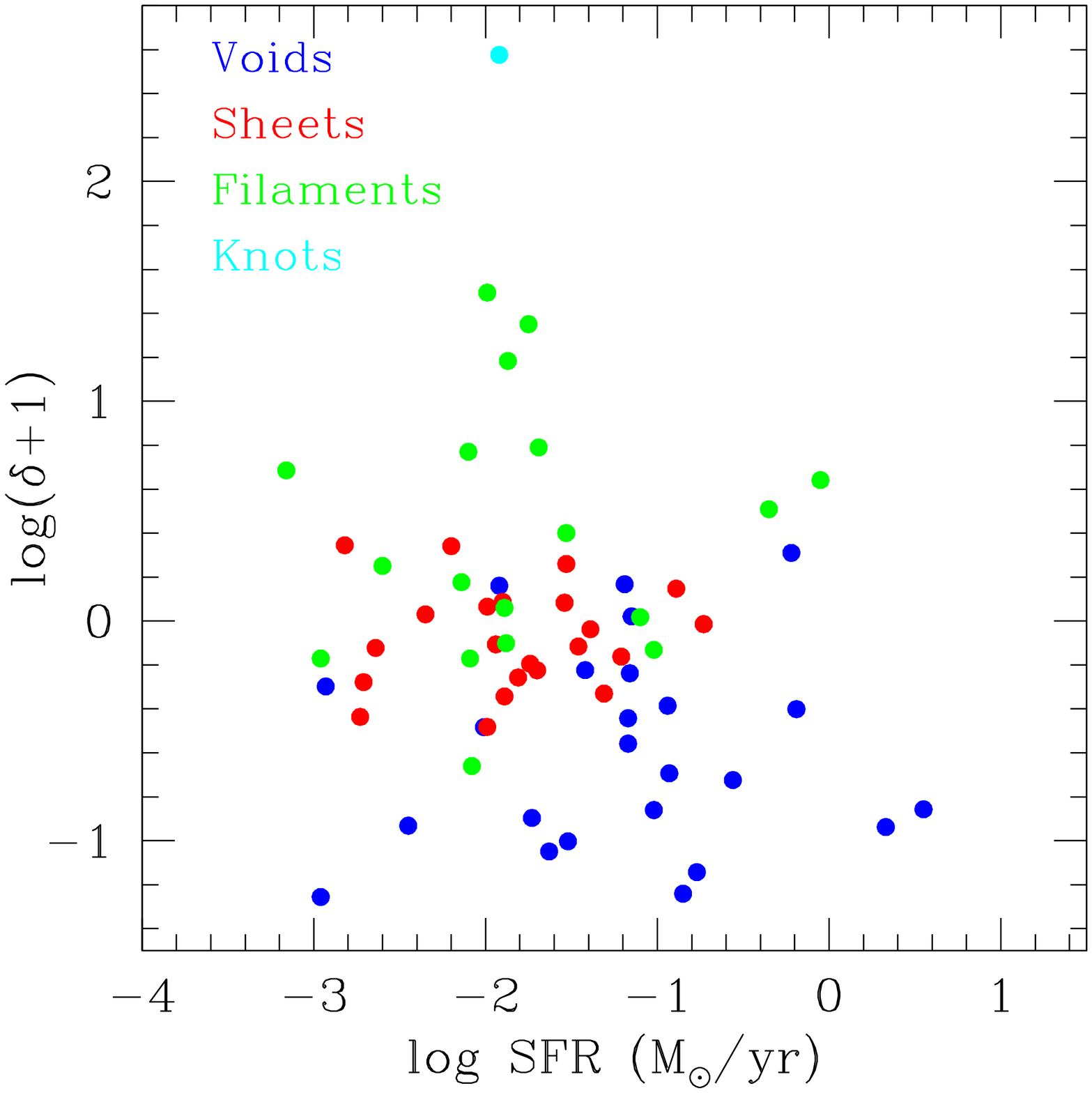}
(b)
\includegraphics[width=6cm]{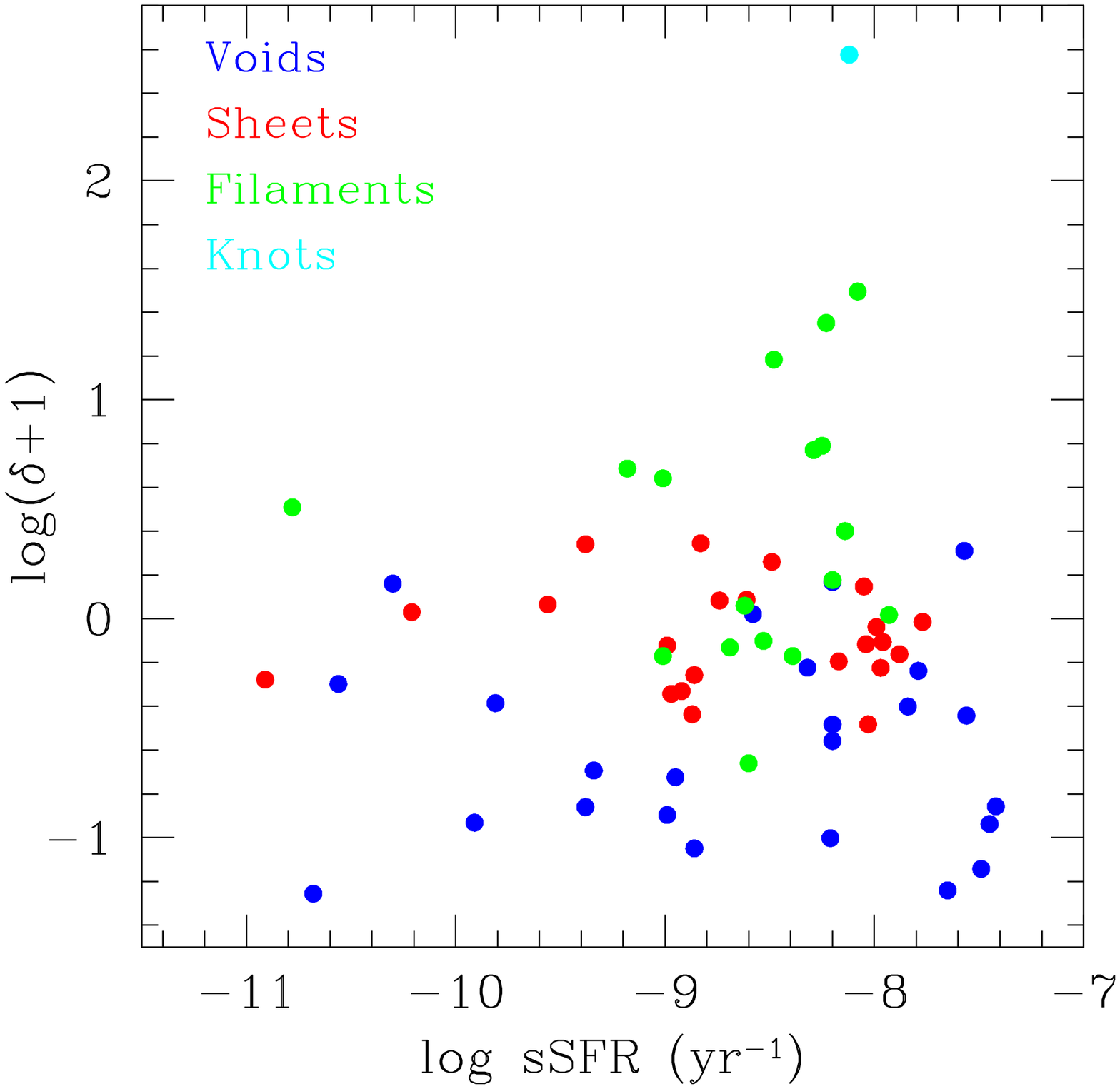}
(c)
\includegraphics[width=6cm]{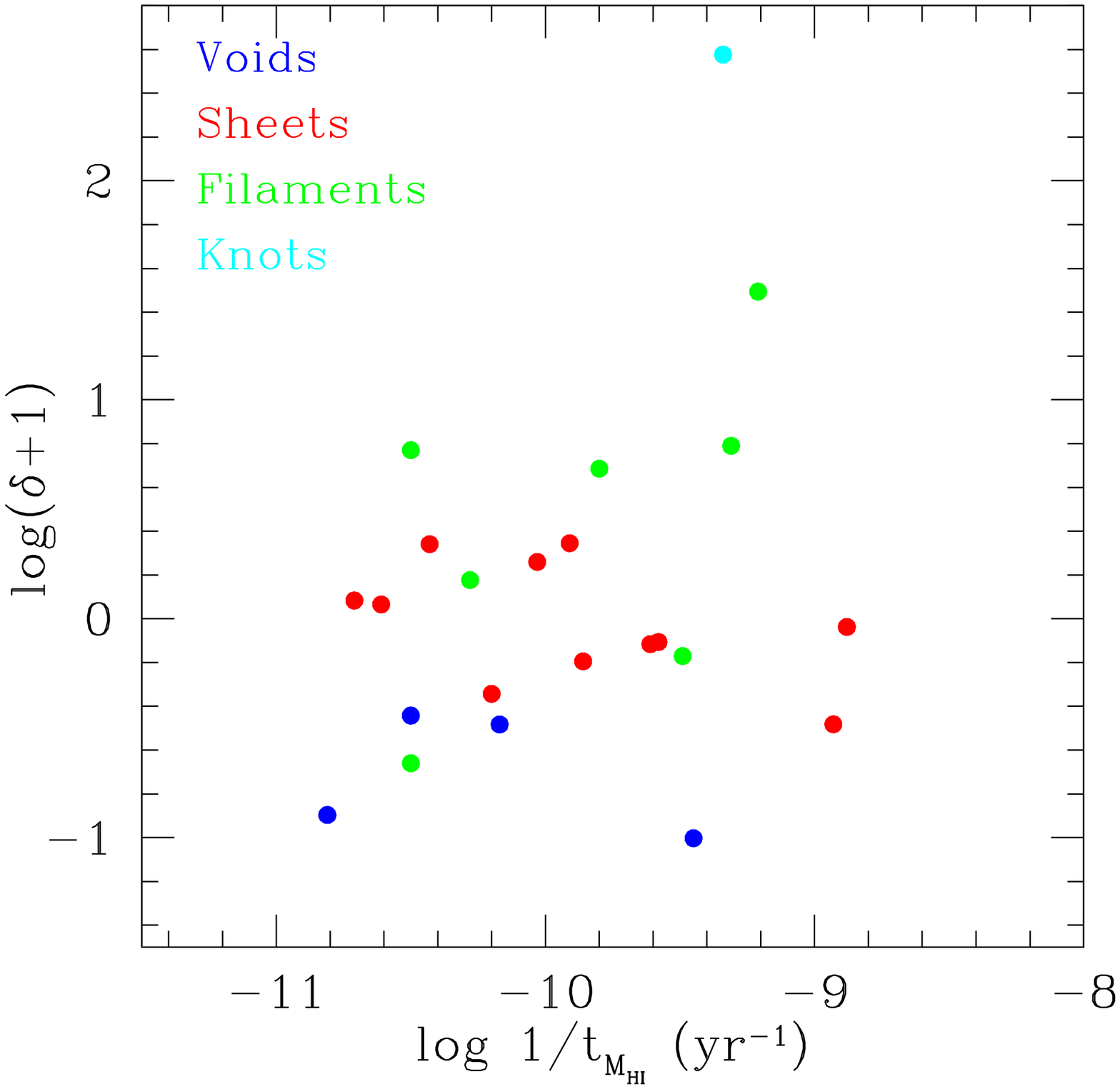}
(d) 
\includegraphics[width=6cm]{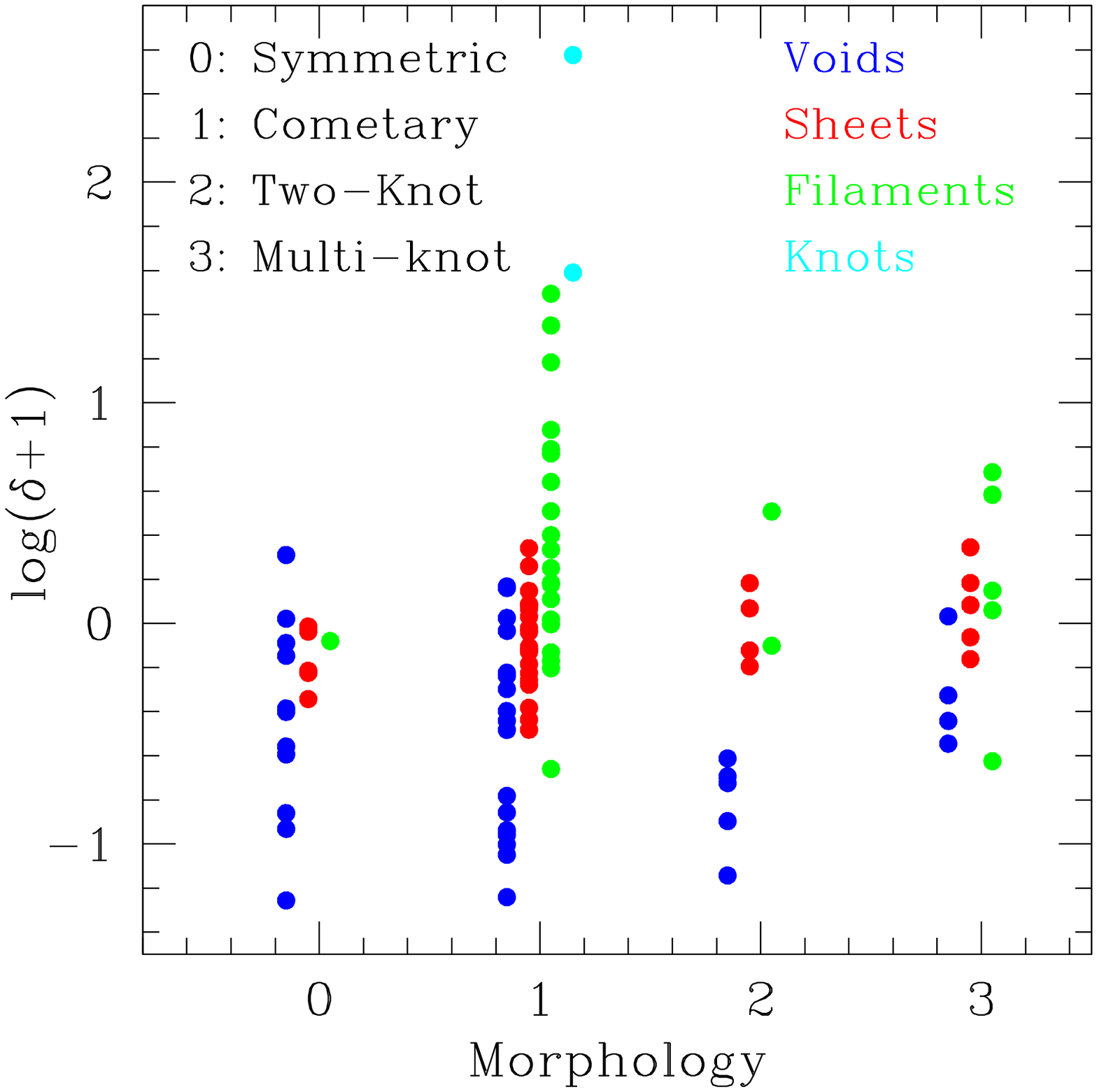}
(e)
\includegraphics[width=6cm]{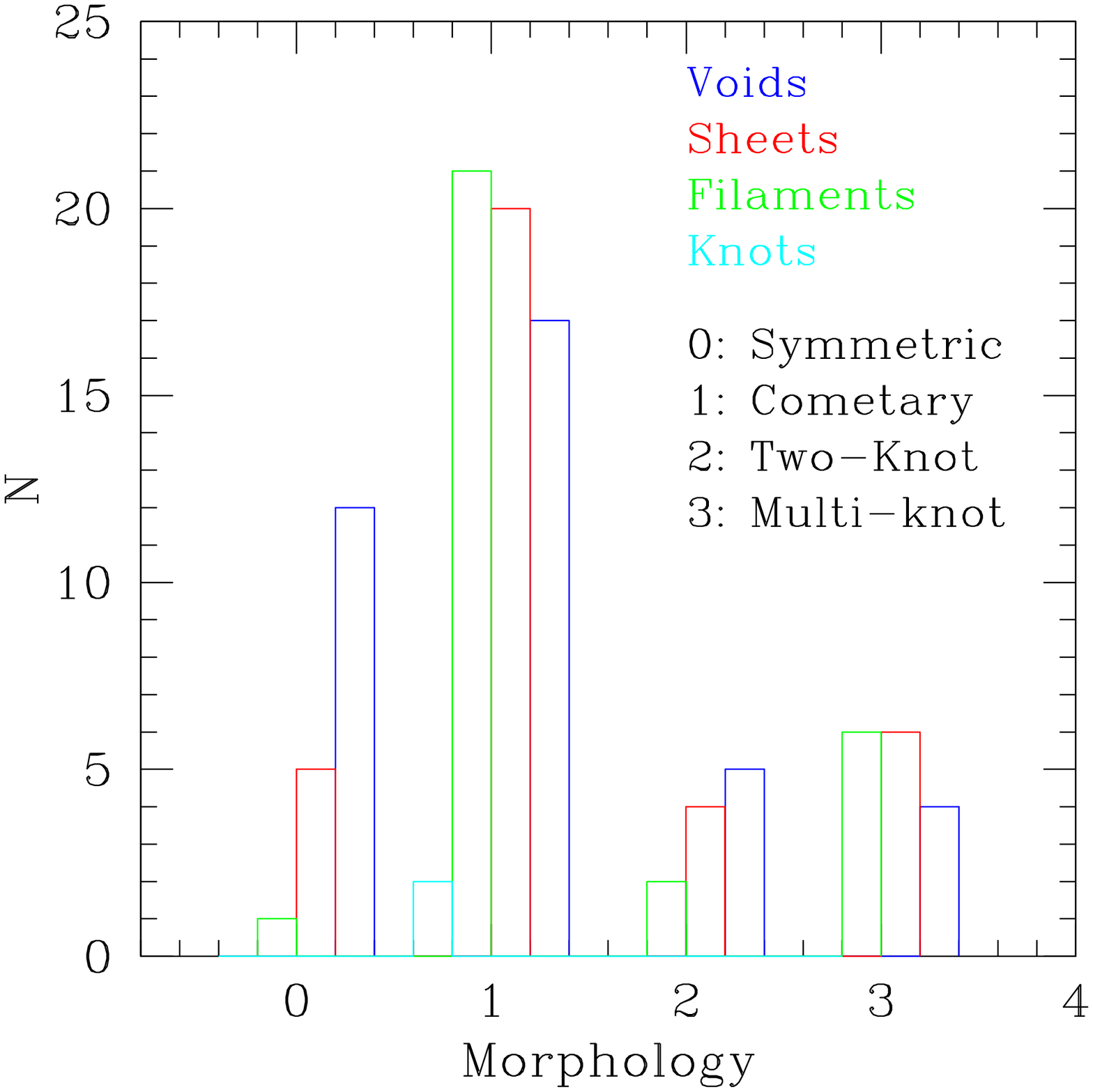}
(f)
\caption{The overdensity as a function of (a) metallicity, (b) SFR, (c) sSFR, (d) \ion{H}{1} gas consumption rate (SFR/M$_{\rm HI}$), and (e and f) optical morphology (symmetric, cometary, two-knot and multi-knot), color-coded by cosmic web type (blue for voids, red for sheets, green for filaments and cyan for knots). These parameters were taken from Table~3 and 4 of Paper~I.}
\end{center} 
\end{figure*} 

%%%%%%%%%%%%%%%%%%%%%%%%%%%

In summary, we find that XMPs tend to reside in low-density environments, and in cosmic web types characterized as voids and sheets. We further find that XMPs, BCDs and Irr galaxies tend to inhabit the same type of environment, which is in stark contrast with the behaviour of E, S0, Sp galaxies in the zero redshift 2MRS sample.

\subsection{Proximity to SDSS Morphological Filaments}

We have investigated the proximity (or inclusion) of XMPs to morphological filaments, i.e., "chains" of galaxies, as empirically defined using galaxies in the spectroscopic sample of the SDSS DR8 dataset. Tempel et al. (2014) use a marked-point process with interactions, called the Bisous model (Stoica, Gregori \& Mateu 2005), to detect galaxy filaments in the spectroscopic sample of the SDSS DR8 dataset (Tempel, Tago \& Liivam\"agi 2012). A redshift interval of 0.009 $ \leq z \leq  $ 0.155 was used so that the lower redshift limit excludes the Local Supercluster. The authors look for morphological filaments with a radius of 0.5~$h^{-1}$~Mpc. They find that the shortest filaments in their filament catalog are 3 -- 5~$h^{-1}$~Mpc in length, while typical filament lengths are of 60~$h^{-1}$~Mpc.

We have mined the catalog of galaxies (hereinafter A3), used to generate the morphological filaments, to search for XMP identifications, or, alternatively, to find the nearest galaxy to each XMP. Distances between the XMPs and A3 galaxies were estimated neglecting galaxy proper motions, since the local flow for pairs of nearby galaxies should be of the same order. A3 also includes the distance of each galaxy to the nearest morphological filament (Tempel et al. 2014). Hence, the distance of the nearest galaxy and morphological filament can be used as a proxy for the distance of the XMPs to morphological filaments. In order to identify our sample XMPs in the A3, we searched the A3 for galaxies within $D \sim$ 10~kpc of the XMP positions; none were found. We are, therefore, confident that the nearest galaxy identified for each XMP is truly a separate galaxy. We stress that A3 contains $\sim$ 500 000 SDSS DR8 sources in a contiguous field, which could miss the XMPs and some of their possible neighbours (see below).

%Because typically these galaxy filaments have a width of 0.5$h^{-1}$~Mpc, we consider $D$ = 0.1~Mpc as a limit to the distance to the nearest galaxy and to the filament axis/end-point (of the nearest galaxy; Fig.~7a and c black line); i.e, if the XMP is within $D$ = 0.1~Mpc of the nearest galaxy, which in turn is within $D$ = 0.1~Mpc of the nearest filament axis/end-point, then we consider that the XMP is also at a small distance ($D \lesssim $ 0.1~Mpc) to the filament.

Figure~6 contains the XMP distance to the nearest galaxy ($D_{\rm galaxy}$) in A3, as a function of the distance between the nearest galaxy in A3 and the nearest morphological filament ($D_{\rm filament}$), color-coded by cosmic web type of the XMP. The figure also contains solid black lines, indicating the radii of the morphological filaments, so that XMPs in galaxy filaments should display $D_{\rm galaxy} \lesssim $ 0.5~Mpc and $D_{\rm filament} \lesssim $ 0.5~Mpc. Only three of the XMPs are found within the $D_{\rm galaxy} \lesssim $ 0.5~Mpc and $D_{\rm filament} \lesssim $ 0.5~Mpc region, and only marginally (Fig.~6). Therefore, we find that XMPs are generally not contained within the morphological filaments defined by Tempel et al. (2014).

%%%%%%%%%%%%%%%%%%%%%%%%%

% Figure 6 - Comparison of XMPs with SDSS galaxies and SDSS galaxy filaments

\setcounter{figure}{5}

\begin{figure}
\begin{center}
\includegraphics[width=6cm]{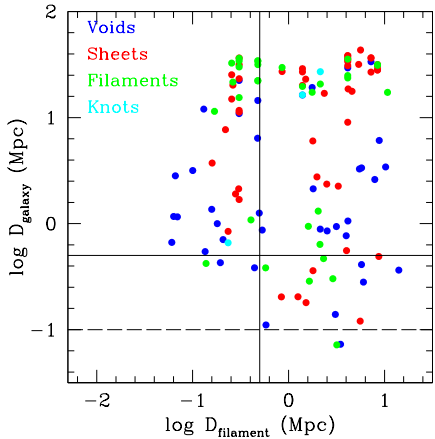} 
\caption{The distance between the XMP and the nearest galaxy in A3, as a function of the distance between the nearest galaxy in A3 and the nearest morphological filament, color-coded by cosmic web type of the XMP. The solid black lines indicate the radii of the morphological filaments, so that XMPs in galaxy filaments should display $D_{\rm galaxy} \lesssim $ 0.5~Mpc and $D_{\rm filament} \lesssim $ 0.5~Mpc. The black dashed line indicates a representative distance for the nearest galaxy, in order to exert some gravitational influence over the XMP (see Sect.~3.3 and 3.4).}
\end{center} 
\end{figure}

\subsection{Nearest Galaxies and Potential Perturbers}

According to the galaxies in the A3 catalog, only two XMPs have a galaxy within $D_{\rm galaxy} \lesssim $ 100~kpc, with one of these XMPs residing in a filament and another in a void (Fig.~6). Indeed, there are four documented XMPs that are known to be interacting with nearby companions (IZw 18, SBS 0335-052, HS 0822+03542 and SBS 1129+576; Sect.~2 and Appendix). 

Overall, the nearest galaxies in A3 show a large range in distances to the XMPs ($D_{\rm galaxy} \sim $ 0.1 -- 100~Mpc), which does not seem to depend strongly on the XMP environment (Fig.~6). However, there is a tendency for XMPs in filaments to show a somewhat bimodal behaviour: there is a group of (nearest) galaxies at smaller ($D_{\rm galaxy} \lesssim $ 1~Mpc) and larger ($D_{\rm galaxy} \gtrsim $ 10~Mpc) distances (Fig.~6).

In order to complement the results based on A3, we have also looked for nearest galaxy neighbours, as potential perturbers, using the full SDSS DR9 (Ahn et al. 2012) catalog. The search was performed within a 12\arcmin~radius around each XMP, corresponding to a distance of $D \sim $ 200~kpc at $z$ = 0.01. We further constrained the galaxies to have spectroscopic redshifts of $z < $ 0.1, resulting in $\sim $1 000 galaxies, potential perturbers to 104 XMPs. We have also estimated the stellar mass of the potential perturbers from their SDSS DR9 $g-r$ colours, using the mass-to-light ratios from Bell \& de Jong (2001).

%We have identified the XMPs in the SDSS DR9 dataset by conservatively constraining the position offset (between the XMP position and the SDSS DR9 source position) to $\Delta \theta \lesssim $ 5\arcsec~($D \lesssim $ 2~kpc at $z \sim $ 0.01). The XMPs identified in this way all exhibit log (M$_{\ast}$/M$_{\odot}$) $\lesssim $ 9 (Fig.~7).

In Figure~7 we plot the stellar mass (M$_{\ast}$) and distance ($D_{\rm perturber}$) of the nearest potential perturber within $D \lesssim $ 300~Mpc of the XMPs, color-coded by cosmic web type of the XMP. Only one XMP shows a potential perturber within $D_{\rm perturber} \lesssim $ 100~kpc, with a stellar mass of M$_{\ast} \sim $ 10$^{7}$~M$_{\odot}$ (Fig.~7). We can not, however, exclude the presence of potential perturbers with no measured spectroscopic redshift in the SDSS, nor less luminous (and less massive) than the XMPs, due to the magnitude cuttoff of the SDSS (magnitude in $r \gtrsim $ 17.7~mag). Comparing Fig.~6 and Fig.~7, we conclude that the potential biases in the A3 sample do not modify the conclusion that XMPs generally do not have close companion galaxies. Above a $D_{\rm perturber} \sim $ 1~Mpc distance to the XMP, the stellar masses of the nearest potential perturbers show a change; the stellar masses not only span a larger range, but potential perturbers with stellar masses of up to M$_{\ast} \sim $ 10$^{11}$~M$_{\odot}$ can be found (Fig.~7). The nearest, least massive, and most massive potential perturbers belong to XMPs in all three major cosmic web types (voids, sheets and filaments). We further note a break in perturber mass distribution at $D_{\rm perturber} \sim $ 10~Mpc; above this distance, stellar masses are skewed towards the larger range (M$_{\ast} \gtrsim $ 10$^{9}$~M$_{\odot}$), consistent with the detection, at larger distances, of more luminous, and hence, more massive galaxies (Fig.~7). 

%%%%%%%%%%%%%%%%%%%%%%%%%

% Figure 7 - Comparison of XMPs with SDSS galaxies 

\setcounter{figure}{6}

\begin{figure}
\begin{center}
%\includegraphics[width=6cm]{/home/mfilho/WORK/NEW_XMPs/Environment/SDSS_spz_12arcmin_Mass_Dist.ps} 
%(a)
%\includegraphics[width=6cm]{/home/mfilho/WORK/NEW_XMPs/Environment/SDSS_spz_12arcmin_Mass_Dist_NO_XMP.ps} 
%(b) \\
\includegraphics[width=6cm]{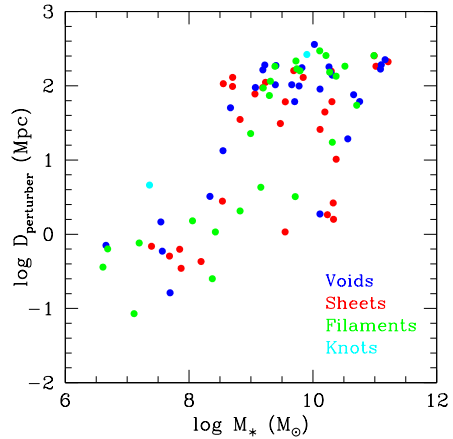}
%(c)
\caption{The distance to the nearest potential perturber in the SDSS DR9 (within $D \lesssim $ 300~Mpc of the XMPs), as a function of stellar mass, color-coded by the cosmic web type of the XMP (blue for voids, red for sheets, green for filaments and cyan for knots). The stellar mass of the nearest potential perturber has been estimated from the SDSS DR9 $g-r$ colours, using the mass-to-light ratios from Bell \& de Jong (2001).}
\end{center} 
\end{figure}

%%%%%%%%%%%%%%%%%%%%%%%%%%%

% There are six XMPs which show potential perturbers within $D \lesssim $ 100~kpc. Three of these nearby potential perturbers are of XMPs in voids, one of an XMP in a sheet, one of an XMP in a filament and one of an XMP in a knot. Moreover, the potential perturbers all have M$_{\ast} \lesssim $ 10$^{8.5}$~M$_{\odot}$, i.e., they are low-mass galaxies, similar to the XMPs themselves. There are four confirmed cases (IZw 18, SBS 0335-053W and SBS 0335-053E, HS 0822+03542 and SBS 1129+576; Section~2) of XMPs with interacting companions, within this distance radius. However, we caution that the SDSS automated pipeline is known to identify multiple sources within one XMP, particularly when there is more than one bright star-forming knot. Indeed, when only the nearest potential perturber is plotted for each XMP (Fig.~8c), 

Overall, with the exception of the already documented cases of XMPs with companions (Sect.~2 and Appendix), statistically we do not find evidence for the presence of potential perturbers (with measured spectroscopic redshifts in the SDSS) within $D \lesssim $ 100~kpc of the XMPs. We caution, however, that due to a lack of spectroscopic redshift and the spectroscopic flux limit of the SDSS, we may be missing potential perturbers of very low mass and luminosity.

\subsection{Perturbations and Interactions}

Because the nearest galaxies to XMPs are generally found to be at distances $D \gtrsim $ 100~kpc (Fig.~6 and 7), we have attempted to evaluate the gravitational influence these could have on the XMPs, particularly if they could potentially trigger and/or feed the star formation. As a first approximation, we have analyzed the conditions for perturbation in a system of a satellite galaxy (XMP) on a circular orbit around a perturber. Let us consider that the satellite galaxy has mass $m$ and radius $r$, orbiting a perturber galaxy of mass $M$, at a distance of $D$, and where $m << M$.

%As a first approximation, we have analyzed the conditions for perturbation of the XMPs using tidal force criteria. 

Minor mergers/interactions (when $m << M$) occur through dynamical friction, when the perturber and satellite share a common dark matter halo. This dynamical friction causes orbital decay of the satellite galaxy within the so-called dynamical friction timescale (e.g.,~Patton et al. 2000; Binney \& Tremaine 2008):

\begin{equation}
%T_{\rm fric} \, {\rm (Gyr)} \simeq \frac{2.64 \times 10^5 \, D^2 \, v}{m {\rm \, ln \; \Lambda}},
T_{\rm fric}\simeq \,2.64 \cdot 10^5 \,{\rm Gyr} \times \frac{D^2 \, v}{m \ln \; \Lambda},
\end{equation}

\noindent where $D$ is the distance between the satellite galaxy and the perturber (in kpc), $v$ is the circular speed of the satellite (in km s$^{-1}$), $m$ is the mass of the satellite (in M$_{\odot}$), and ln $\Lambda$ is the Coulomb logarithm. This expression does not take into consideration any mass loss (from tidal stripping) that the satellite may suffer during the orbital decay.

If we assume a typical distance to the perturber of $D \simeq $ 100~kpc (Fig.~6 and 7), a satellite mass of $m \simeq $ 10$^8$~M$_{\odot}$ (Fig.~4c), a Coulomb logarithm of 2 (e.g.,~Binney \& Tremaine 2008; Filho et al. 2014), and a circular velocity of the satellite of $v \simeq \sqrt {G \, M / D}$, where $G$ is the gravitational constant, this then yields a dynamical friction timescale of $T_{\rm fric} \simeq $ 50~Gyr, for $M \simeq$ m, or 150~Gyr, for $M \simeq 10 \, m$. This estimate provides a timescale for the satellite to spiral in towards the perturber. The resulting timescales are too long for the merger/interaction process to be significant in XMPs.

% M = 10^8 -> v = 3 km/s
% M = 10^9 -> 7 km/s
% v = 282.1777693 m/s
% v = 6659.05 m/s = 6.55905 km/s

As the satellite approaches the perturber on this orbital decay, the tidal forces at work on the satellite become more important. If we now consider a test particle at a distance $\Delta r$ from the surface of the satellite (where $\Delta r << r$), then the perturber will exert a perturbative tidal force on the test particle (relative to the force exerted at the center of the satellite), that is given by (e.g.,~Binney \& Tremaine 2008):

\begin{equation}
F_{\rm M} \simeq \frac{2 \, G \, M \, (r + \Delta r)}{D^3} \simeq \frac{2 \, G \, M \, r}{D^3}.
\end{equation}

\noindent The satellite will also exert a restoring force on the test particle, if tidal forces displace it from its equilibrium configuration at $r$, namely, (e.g.,~Binney \& Tremaine 2008):

%a force on the test particle (gravitational binding force), which is given by:

\begin{equation}
F_{\rm m} = \frac{G \, m \, \Delta r}{r^3},
\end{equation}

\noindent where we have neglected the pressure support in the satellite (Adams et al. 2011). In equilibrium $F_{\rm M} \simeq F_{\rm m}$; therefore, the displacement produced by tidal forces is given by:

\begin{equation}
\frac{\Delta r}{r} \simeq 2 \, \frac{M}{m} \left( \frac{r}{D} \right) ^3.
\end{equation}

\noindent The condition for tidal disruption of the test particle by the perturber is producing a large perturbation, i.e., $\Delta r/r \gtrsim $ 1.

%Therefore, the condition for disruption of the test particle by the perturber is that the perturbative (tidal) force must be larger than the gravitational binding force, $F_{\rm M} > F_{\rm m}$. This criteria is similar to the criteria applied for tidal stripping used in cluster, merger and binary studies, but where $\frac{\Delta r}{r} \simeq 1$.

In Figure~8 we show the displacement produced by tidal forces (Eq.~4), as a function of the ratio between the XMP radius and the distance to the perturber, color-coded by the relative XMP/perturber mass. As can be seen from Figure~8, in order for tidal stripping to occur, the perturber is required to have a high mass and/or be located at a small distance to the satellite. For a perturber and satellite of similar mass ($M \simeq  m$), the distance must be of the order of the size of the galaxy ($D \simeq r$). Because XMPs have radii of a few kpc, $D$ must also be a few kpc in order to exert appreciable gravitational influence on the XMP.

%the "normalized" (by gravitational constant and test particle mass) perturbative "force" (Eq.~2) as a function of perturber mass (M) and magnitude of the perturbation ($\Delta r$), for a satellite of mass m = 10$^{9}$ M$_{\odot}$ and radius $r =$1~kpc, color-coded by the distance to the perturber ($D$). The black lines in the plots show the "normalized" gravitational binding "force" for a satellite of mass m = 10$^{9}$ M$_{\odot}$, radius $r =$1~kpc (Eq.~3) and magnitude of perturbation of $\Delta r = $0.001, 0.01, 0.1, and 1~kpc, or 0.01, 0.1,  1 and 100\% of the satellite radius $r$.

%%%%%%%%%%%%%%%%%%%%%%%%%

% Figure 8 - Tidal/Perturbative Force

\setcounter{figure}{7}

\begin{figure}
\begin{center}
\includegraphics[width=6cm]{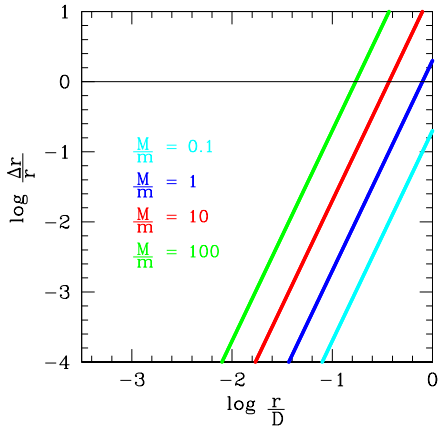}
\caption{The displacement produced by tidal forces, as a function of the ratio between the XMP radius and the distance to the perturber, color-coded by the relative XMP/perturber mass: $M/m$ = 0.1 (cyan), 1 (blue), 10 (red), 100 (green). The black line corresponds to $\Delta r/r$ = 1, the lower limit criteria for tidal stripping.}
\end{center}
\end{figure}

%%%%%%%%%%%%%%%%%%%%%%%%%%%

Therefore, although a significant number of XMPs show potential perturbers within $D \sim $ 300~Mpc, it is unlikely that these will cause major gravitational disturbance (or tidal stripping) of the XMP, which could instigate gas transfer and sustain star formation.

\section{Discussion and Conclusions}

XMPs are found to tendentiously reside in low-density environments ($\sim$ 60\%) and in cosmic web types characterized as voids and sheets ($\sim$ 75\%; Sect.~3.1). In this respect, their behaviour is extreme, in contrast to the location in the cosmic web of E, S0 and Sp galaxies, but inhabiting regions similar to those of Irrs and BCDs (Sect.~3.1). 
 
We find results similar to those found for nearby (relatively) luminous galaxies (Blanton \& Moustakas 2009) and H$\alpha $-emitting galaxies (Darvish et al. 2014), whereby the environment does not determine the overall observed properties (mass, luminosity, SFR, etc.) of a certain galaxy type. However, the fraction of a certain galaxy type (in this case, XMPs), does appear to be determined by the environment in which it resides (Sect.~3.1).

%With the exception of IZw 18 (Lelli et al. 2012a; Lelli et al. 2014a, Lelli et al. 2014b), SBS 0335-052W, SBS 0335-052E (Ekta, Pustilnik \& Chengalur 2009), SBS 1129+576  (Ekta, Chengalur \& Pustilnik 2006; all of which have been confirmed with \ion{H}{1} data to be interacting with companion galaxies), and possibly Sag Dig (potential merger between dwarfs; Beccari et al. 2014), we find no statistical evidence in the XMP population for (major) interactions with (gas-rich) galaxy companions in the recent past (Section~2 and 3), a mechanism which is classically evoked as a driver for star formation and overall galaxy evolution (e.g.,~Keel et al. 1985; e.g.,~Robaina et al. 2009). If Sag Dig is the result of a recent merger between dwarf galaxies, as is suggested in Beccari et al. (2014), then the merger signatures present in Sag Dig (clumpy \ion{H}{1} distribution, the large offset between the optical center and \ion{H}{1} distribution, no clear rotational motion) are generally absent in the XMP population. 

The evidence for occupation of low-density environments is further strengthened when we investigate the proximity to morphological filaments (Sect.~3.2), and galaxies in the vicinity of the XMPs (Sect.~3.3). We find no significant evidence for the presence of galaxy filaments within $D \sim $ 0.5~Mpc of an XMP, and only one potential perturber (of mass and luminosity at least equal to that of an XMP, and with a measured spectroscopic redshift in the SDSS) is found within $D \sim $ 100~kpc of an XMP. However, even at these distances, we have demonstrated that a potential perturber is very unlikely to cause gravitational disturbance capable of drawing fresh gas or driving the star formation (Sect.~3.4).

We have also analyzed the small-scale environment, probing for signatures of galaxy merger/interactions in the published \ion{H}{1} data of XMPs (Sect.~2 and Appendix). IZw 18 (Lelli et al. 2012a; Lelli, Verheijan \& Fraternali 2014a, b), SBS 0335-052W and SBS 0335-052E (Ekta, Pustilnik \& Chengalur 2009), HS 0822+03542 (Chengalur et al. 2006) and SBS 1129+576  (Ekta, Chengalur \& Pustilnik 2006) have been confirmed to be interacting with companion galaxies, all within $D \lesssim $ 25~kpc (Sect.~2 and Appendix). Moreover, it has been suggested that the \ion{H}{1} properties observed in Sag Dig could be the result of a recent merger between dwarf galaxies (Beccari et al. 2014). However, if Sag Dig is the result of a merger (Beccari et al. 2014; Sect.~2 and Appendix), then the merger signatures present in Sag Dig (ring-like \ion{H}{1} distribution versus disk-like optical morphology, clumpy \ion{H}{1} distribution with no optical correspondence, large offset between the optical center and the depression in the \ion{H}{1} distribution, no clear rotational motion) are generally absent in the XMP population. Indeed, we find no statistical evidence for (major) interactions with (gas-rich) galaxy companions in the recent past (Sect.~2 and Appendix; see also Brosch, Almoznino \& Heller 2004, Holwerda et al. 2013 and Kreckel et al. 2015), a mechanism which is traditionally invoked as a driver for star formation and overall galaxy evolution (e.g.,~Keel et al. 1985; Robaina et al. 2009). Although such an interaction could provide large quantities of fresh \ion{H}{1} gas, and create conditions for triggering/feeding the observed star formation, it can not explain why the gas we observe is metal-poor (Paper~I), nor can it explain the metallicity inhomogeneities observed in some XMPs (S\'anchez Almeida et al. 2013, 2014a), and it does not explain why we do not systematically observe signatures of major disturbances/interactions in the \ion{H}{1} gas, and particularly in the \ion{H}{1}-to-optical morphological/kinematical relation (Sect.~2 and Appendix). In fact, the central regions of the \ion{H}{1} structures are relatively well-behaved; in many cases they are organized in a disk with clear rotational motion, in which the kinematical \ion{H}{1} axis is virtually aligned (within 15$^\circ$) with the \ion{H}{1} and optical axis, where the peak of the \ion{H}{1} emission is positionally coincident (within 10\arcsec) with the center of the optical disk or the brightest star formation knot (Appendix), and where the velocity offsets between the stellar and \ion{H}{1} components are absent or small ($<$ 40 km s$^{-1}$; of the order of the velocity dispersion; Paper~I). Hints of kinematical and morphological disturbance of the \ion{H}{1} gas appear primarily in the galaxy outskirts; tails, extensions, clumps of \ion{H}{1} gas, and kinks, discontinuities, twisting of the velocity field (Sect.~2 and Appendix). 

Indeed using three different diagnostic tools (overdensity, cosmic web type and nearest galaxies/potential perturbers), we find evidence that XMPs are relatively isolated in the nearby Universe, on scales of hundreds of kpc (Sect.~2 and 3). However, despite this isolation, we must understand how XMPs acquire such large amounts of metal-poor \ion{H}{1} gas (Paper~I and references therein), and how they sustain their star formation.

An alternative scenario is that XMPs are truly primordial galaxies, in the sense that they have remained quiescent for some time, thus preserving the \ion{H}{1} gas acquired early on in their formation. However, this scenario can not explain the current starburst phase of the XMPs, nor can it explain the reason, or provide a mechanism, for the XMPs to have abandoned their quiescent phase. Moreover, because most XMPs have huge SFRs for their stellar masses (Fig.~5c), they would have had to form all their stars in less than a Gyr, at the present rate.

Following S\'anchez-Almeida et al. (2013, 2014b), we find that the XMP properties, particularly their environment characteristics, are consistent with XMP evolution being mainly driven by metal-poor gas accretion, similar to the cold accretion flows simulated at high redshift (Kere\v{s} et al. 2005, 2009; Dekel et al. 2009; Dekel, Sari \& Ceverino 2009; Brooks et al. 2009; Ceverino, Dekel \& Bournaud 2010; Ceverino et al. 2014; Nuza et al. 2014a), and supported by observational evidence (Cresci et al. 2010; van de Voort et al. 2012). Such accretion is known to be stochastic, occurring along streams of high density cold gas, which are capable of penetrating low-mass haloes (Kere\v{s} et al. 2005, 2009; Dekel et al. 2009; Dekel, Sari \& Ceverino 2009; Brooks et al. 2009; Verbeke et al. 2014). Moreover, this cold-mode  of accretion is able to explain the major observational properties of the XMPs: the relative isolation (Sect.~2 and 3), the lack of recent interaction/merger signatures in the gas and stars (Sect.~2 and Appendix), the lopsided optical morphology (Papaderos et al. 2008; Morales-Luis et al. 2011; Paper~I), the large amount of metal-poor \ion{H}{1} gas (Paper~I and references therein), the metallicity inhomogeneities (S\'anchez Almeida 2013, 2014a), and the recent burst of star formation.

\begin{acknowledgements}

Thanks are due to Debra and Bruce Elmegreen for insightful discussions on the nature of XMPs, to Daniel Ceverino and Claudio Della Vecchia for advice with the estimates in Section~3.4, and to Ricardo Amor\'\i n for the estimation of XMP diameters given in Table~1. We would like to thank the anonymous referee for very pertinent comments that have greatly improved this manuscript. Research by M.E.F. is supported by the Canary Islands CIE: Tri-Continental Atlantic Campus. This research has been partly funded by the Spanish Ministery of Economy, project $\mathrm{AYA}2013-47742-\mathrm{C}4-2-\mathrm{P}$. S.E.N. acknowledges support by the Deutsche Forschungsgemeinschaft under the grants $\mathrm{NU}332/2-1$ and $\mathrm{MU}1020 16-1$. S.H. acknowledges support by the Deutsche Forschungsgemeinschaft under the grant $\mathrm{GO}563/21-1$. The constrained simulations have been performed at the Juelich Supercomputing Centre (JSC). This research has made use of the NASA/IPAC Extragalactic Database (NED), the Digital Sky Survey (DSS), and the Sloan Digital Sky Survey (SDSS) Data Release (DR) 6 -- 10.

\end{acknowledgements}

%%%%%%%%%%%%%%%%%%%%%%%%%
%\bibliographystyle{aa}%

%\begin{thebibliography}

%\end{thebibliography}

\begin{appendix}

\section{Description of XMPs With Archival High Resolution \ion{H}{1} Observations}

This appendix provides a detailed description of the \ion{H}{1} properties of the 19 XMPs with high resolution interferometric data in literature. The average properties of these galaxies, as a class, are discussed in Section~2. Here we describe the peculiarities of the individual sources used to distill a global picture. We cite the original references of the archival data used in our analysis, as well as the main observational properties of the targets inferred from the visual inspection of the published maps.

Our description of the high resolution \ion{H}{1} properties of the XMPs are based on the definition of several source parameters: the position angle of the largest \ion{H}{1} extension, measured counterclockwise from the West (PA$_{\rm HI}$), the position angle of the largest optical extension, measured counterclockwise from the West, from the SDSS or DSS images (PA$_{\rm opt}$), the position angle of the \ion{H}{1} velocity gradient, measured counterclockwise from the West (PA$_{\rm VG}$), the position angle of the \ion{H}{1} velocity field (PA$_{\rm kin}$), the offset between the \ion{H}{1} peak and/or the optical center/brightest star-forming knot ($\Delta D$), and the offset between the \ion{H}{1} velocity gradient or largest extension position angle and the optical largest extension position angle ($\Delta \theta)$. References and survey/interferometer information (GMRT -- Giant Metrewave Telescope, VLA -- Very Large Array, THINGS -- The \ion{H}{1} Nearby Galaxy Survey, ACS - Advanced Camera for Surveys, ANGST -- ACS Nearby Galaxy Survey Treasury) for the \ion{H}{1} data are provided for each of the sources.

{\bf J0119-935}	From Ekta \& Chengalur (2010; their Fig.~3 and 4), with the GMRT. The lower resolution (46\arcsec~x 32\arcsec) \ion{H}{1} map shows a single-peaked "boomerang-type" Northeast-Southwest disk structure (PA$_{\rm HI} \approx $ 120$^\circ$), with an extension to the South and to the North, the latter which connects to a clump of emission to the Northwest. The orientation of the \ion{H}{1} disk is roughly coincident with the position angle of the optical cometary emission (PA$_{\rm opt} \approx $ 115$^\circ$). However, the center of the \ion{H}{1} emission does not coincide with either the center of the optical disk, nor the bright star formation knot that leads the cometary structure in the Northeast ($\Delta D \lesssim $ 10\arcsec). In the central regions, the \ion{H}{1} velocity gradient (PA$_{\rm VG} \approx $ 120$^\circ$) follows the \ion{H}{1} and optical morphology. However, in the outskirts, the \ion{H}{1} velocity field displays distortions, with the mimicking of the "boomerang" shape observed in the \ion{H}{1} morphology. The intermediate resolution (30\arcsec~x 20\arcsec) image shows a "Y-shaped" \ion{H}{1} morphology, cradling the bright optical star formation knot to the Northeast, with the Southern extension curving towards the East. The position angle at this resolution (PA$_{\rm HI} \approx $ 110$^\circ$) is roughly coincident with the larger-scale \ion{H}{1} and optical morphology. 
%{\it \ion{H}{1}/opt $\sim$ 4; Type A/B large scale and Type B small scale}

{\bf UM 133} From Ekta \& Chengalur (2010; their Fig.~1 and 2), with the GMRT. The lower resolution (42\arcsec~x 36\arcsec) map of the \ion{H}{1} gas displays a single-peaked Northeast-Southwest disk (PA$_{\rm HI} \approx $ 105$^\circ$), with a small extension to the South and a larger extension to the East. The \ion{H}{1} and optical coincide in terms of position angle and geometric (disk) center/\ion{H}{1} peak. There is a less than 10\arcsec~distance between the \ion{H}{1} peak and the bright star formation knot to the Southwest, leading the cometary structure. In the central regions, the \ion{H}{1} gas velocity gradient (PA$_{\rm VG} \approx $ 105$^\circ$) is smooth and parallel to both the optical (PA$_{\rm opt} \approx $ 105$^\circ$) and \ion{H}{1} morphology, which is consistent with a rotating disc. However, distortions are seen in the outskirts, particularly towards the South and East, in the direction of the \ion{H}{1} extensions. At an intermediate resolution (22\arcsec~x 17\arcsec), the \ion{H}{1} morphology is "arrow-shaped", pointing towards the Northeast, and accompanying the overall optical structure. Similarly to the lower resolution images, extensions to the South and East are observed. The morphology and central velocity gradient on these scales, match the larger-scale \ion{H}{1} and optical morphology. Although in the central regions the velocity field is well-behaved, in the periphery there are clear asymmetries: the velocity gradient changes to the North-Northeast, and large irregularities are detected to the South-Southeast. 
%{\it \ion{H}{1}/opt $\sim$ 2; Type A/B large scale and Type B small scale}

{\bf SBS 0335-052E and SBS 0335-052W} From Ekta, Chengalur \& Pustilnik (2009; their Fig.~1, 4 and 5), with the GMRT. The low resolution ($\sim $ 40\arcsec) map shows this interacting pair of XMP galaxies ($\Delta D \sim $ 20~kpc), enveloped in a common \ion{H}{1} structure (PA$_{\rm HI} \approx $ 175$^\circ$), with each \ion{H}{1} peak corresponding to a galaxy. At intermediate ($\sim $ 20\arcsec) resolution, the individual single-peaked \ion{H}{1} structures are distinguishable. The elongated disks are oriented along approximately the same direction (PA$_{\rm HI}\approx $ 145$^\circ$ for the Eastern, and PA$_{\rm HI}\approx $ 150$^\circ$ for the Western, component). SBS 0335-052W displays a small nodule of emission to the Southwest. The \ion{H}{1} peaks are slightly offcenter ($\Delta D \lesssim $ 10\arcsec ) from the peaks of optical emission. In the case of SBS 0335-052E, the extended \ion{H}{1} emission is approximately perpendicular to the Northwestern extension seen in the optical image (PA$_{\rm opt} \approx $ 45$^\circ$). The velocity gradient in SBS 0035-052W changes from the center (approximately North-South) to the outskirts (Northeast-Southwest). For SBS 0335-052E, the velocity gradient is along the tidal tail (Northeast-Southwest), but deviates from a standard disk. High resolution ($\sim $ 9\arcsec) observations show that SBS 0335-052W maintains the overall \ion{H}{1} position angle seen at lower resolutions, with an extension to the Northeast. The direction of the velocity gradient is approximately Northwest-Southeast. SBS 0335-052E reveals extended emission (PA$_{\rm HI}\approx $ 20$^\circ$), roughly aligned with the optical cometary tail. 
%{\it East \ion{H}{1}/opt $\sim$ 4; Type B/C} {\it West \ion{H}{1}/opt $\sim$ 6; Type B/C}

{\bf UGC 4305 (also known as Holmberg II, Arp 268, DDO 050 or VII Zw 223)} From Walter et al. (2008; their Fig.~13), from the VLA THINGS. High resolution (6.95\arcsec~x 6.05\arcsec) \ion{H}{1} maps show an almost symmetric spiral-like structure, the internal \ion{H}{1} morphology is complex, displaying holes, clumps and arms. The optical loop appears associated with the brighter \ion{H}{1} spiral arm in the South, which encloses a prominent \ion{H}{1} hole. The Southern (fainter) arm extends as a tail to the Southwest. The \ion{H}{1} velocity gradient runs approximately North-South (PA$_{\rm VG}\approx $ 100$^\circ$), within $\Delta \theta \sim $ 15$^\circ$ of the optical position angle (PA$_{\rm opt}\approx $ 115$^\circ$). 
%{\it \ion{H}{1}/opt $\sim$ 5; Type A/B}

% The \ion{H}{1} morphology and kinematical axis (PA$_{\rm \ion{H}{1}}\sim $PA$_{\rm kin}\approx $100$^\circ$ ) coincide roughly with the optical position angle (PA$_{\rm opt}\approx $115$^\circ$ ). 

{\bf HS 0822+03542} From Chengalur et al. (2006; their Fig.~1, 4 and 5), with the GMRT. The low resolution (42\arcsec~x 37\arcsec) map shows the XMP target along with its pair to the Northeast, the low surface brightness galaxy SAO 0822+545, at a distance of $D \sim$ 11~kpc and a position angle of PA $\approx $ 125$^\circ$. At this resolution, the \ion{H}{1} gas is single-peaked, with an extension to the East-Northeast. The velocity field (27\arcsec~x 23\arcsec) displays a distortion, with the velocity gradient Northwest-Southeast (PA$_{\rm VG} \approx $ 70$^\circ$) near the center, aligned with the \ion{H}{1} morphology (PA$_{\rm HI} \approx $ 70$^\circ$), and approximately East-West (PA$_{\rm VG} \approx $ 170$^\circ$), roughly perpendicular to the \ion{H}{1}, towards the Southeast. Intermediate and high resolution (17\arcsec~x 15\arcsec and 13\arcsec~x 11\arcsec) images show a single-peaked, Northwest-Southeast disk (PA$_{\rm HI} \approx $ 70$^\circ$), with an extension to the Northwest-West, and a clump of emission to the Northeast, the latter tracing the extension seen at lower resolution. The extension to the Northwest-West appears to mimic the position angle of the optical cometary tail (PA$_{\rm opt} \approx $ 45$^\circ$). There is an offset between the \ion{H}{1} peak and the center of the optical disk and bright star formation knot to the Southeast ($\Delta D \lesssim $ 10~arcsec). At the highest resolution (7\arcsec~x 5.5\arcsec), the \ion{H}{1} gas morphology resembles two almost vertical (PA$_{\rm HI} \approx $ 80$^\circ$) "fingers", pointing to the bright star formation knot, and peaking on either side. Overall, there is an offset of $\Delta \theta \sim $ 30$^\circ$ between the optical and \ion{H}{1} alignment. 
%{\it \ion{H}{1}/opt $\gtrsim$ 10 ; Type B/C}

{\bf UGC 4459 (also known as DDO 053 or VII Zw 238)} From Walter et al. (2008; their Fig.~17), from VLA THINGS, and Begum et al. (2006; their Fig.~4), with the GMRT. The \ion{H}{1} map at intermediate resolution (29\arcsec~x 27\arcsec) shows a double-peaked quasi-symmetric spherical structure, with a small extension to the Northeast. The two \ion{H}{1} peaks are roughly coincident with the opposite ends of the optical main disk, with an \ion{H}{1} depression in between. There is slight \ion{H}{1} enhancement associated with the secondary optical disk. The overall position angle of the \ion{H}{1} emission (PA$_{\rm HI} \approx $ 150$^\circ$) is approximately perpendicular to the optical emission (PA$_{\rm opt}\approx $ 55$^\circ$). The velocity gradient direction (PA$_{\rm VG}\approx $ 45$^\circ$) is relatively well-aligned with the optical morphology. The velocity gradient is not uniform, with the field displaying an asymmetry, weighted towards the Southeast. Similarly to the intermediate resolution images, the high resolution (6.34\arcsec~x 5.67\arcsec) \ion{H}{1} maps show two high-density \ion{H}{1} clumps, with a central depressed region, embedded in an \ion{H}{1} halo, with faint Eastern-Northeastern emission. The \ion{H}{1} emission associated with the fainter optical disk is more apparent. On these scales, the velocity gradient direction is roughly aligned with the optical morphology and the larger-scale velocity gradient, and mimics what occurs at low resolution: there are deviations towards the Southeast. 
%{\it \ion{H}{1}/opt $\sim$ 3 Type A/B}

{\bf UGC 4483 (also known as Mrk 0090 or SBS 0826+528)} From Ott et al. (2012; their Fig.~7), from the VLA ANGST, and Lelli et al. (2012b; their Fig.~1 and 2) and Lelli, Verheijan \& Fraternali (2014a, b; their Fig.~1, C.11, and Fig.~1, 2), with the VLA. The intermediate resolution (20\arcsec~x 20\arcsec) \ion{H}{1} map shows almost vertical (PA$_{\rm HI} \approx $ 80$^\circ$) "arrow-shaped" emission, where the head of the arrow spatially coincides with the bright star formation knot to the North, heading the cometary optical morphology. There is also extended \ion{H}{1} emission to the South and Northwest. The central \ion{H}{1} morphology matches well the observed "X-shaped" and overall optical morphology (PA$_{\rm opt} \approx $ 80$^\circ$). The kinematical axis is vertical (PA$_{\rm kin}\approx $ 90$^\circ$), roughly matching the position angle of the brighter arm (PA$_{\rm opt}\approx $ 95$^\circ$) associated with the "X-shaped" optical morphology. The velocity field can be modeled by a rotating disk, possibly with a radial component. High resolution (10\arcsec~x 10\arcsec) \ion{H}{1} observations mimic the larger-scale arrow-shape emission, but two \ion{H}{1} peaks are resolved, the brightest of which is elongated to the Southeast and spatially coincides with the bright optical star formation knot. There is extended emission also to the South and East. On these scales, the \ion{H}{1} major axis (PA$_{\rm HI}\approx $ 60$^\circ$) appears to follow in the direction of the fainter "arm" associated with the "X-shaped" optical morphology (PA$_{\rm opt}\approx $ 60$^\circ$). The \ion{H}{1} kinematical axis is vertical (PA$_{\rm kin}\approx $ 90$^\circ$), and matched to the larger-scale velocity gradient. The highest resolution ($\sim $ 6\arcsec) \ion{H}{1} image is similar to its larger-scale counterparts: the overall morphology is arrow-shaped (PA$_{\rm HI} \approx $ 80$^\circ$), with two resolved \ion{H}{1} peaks, the brightest of which is positionally coincident with the bright star-forming knot. On these scales, the velocity gradient (PA$_{\rm VG}\approx $ 120$^\circ$) is misaligned relative to the \ion{H}{1} and optical (PA$_{\rm opt}\approx $ 80$^\circ$) major axes. The velocity field displays asymmetries and distortions along the axis. 
%{\it \ion{H}{1}/opt $\sim$ 2; Type A}

{\bf IZw 18 (also known as UGCA 166, Mrk 0116 or SBS 0930+554)} From Lelli et al. (2012a; their Fig. 2, 3 and 4) and Lelli, Verheijan \& Fraternali (2014a, b; their Fig.~C.16, and Fig.~1, 2), with the VLA. The intermediate resolution (20\arcsec~x 20\arcsec) \ion{H}{1} image shows single-peaked "triangle-shaped" emission (PA$_{\rm HI} \approx $ 45$^\circ$), with a Southeast extension that curves due South, along $D \sim $ 14~kpc. The main body of the \ion{H}{1} emission is aligned with the optical morphology (PA$_{\rm opt}\approx $ 45$^\circ$). The peak of the \ion{H}{1} coincides with the position of the brightest (Northwest) star formation knot in the optical image. There is small extended emission to the Northeast, North and Northwest. The latter emission is likely associated with the companion galaxy, IZw 18C, with which it is interacting, at a distance of $D \sim $ 18~kpc. The kinematical axis (PA$_{\rm kin}\approx $ 50$^\circ$) is roughly aligned with both the (main) \ion{H}{1} and optical emission. The large Southern extension is at an almost constant velocity. At high resolution (5\arcsec~x 5\arcsec), the \ion{H}{1} emission is resolved into two peaks, positionally coincident with the two brightest star formation knots in the optical image. There is also a halo of extended emission to the Northwest, englobing the interacting companion, IZw 18C. The velocity field (PA$_{\rm kin}\approx $ 55$^\circ$) can be modeled by a rotating disk plus radial motion, with signs of distortion in the direction of IZw 18C. At even high resolution (2\arcsec~x 2\arcsec), there is peaked \ion{H}{1} emission extended to the South/Southeast, in an inverted "V-shaped" morphology, and extended emission to the Northwest, in the direction of IZw 18C. 
%{\it \ion{H}{1}/opt $\sim$ 7; Type A}

{\bf Leo A (also known as Leo III, UGC 5364 or DDO 069)} From Begum et al. (2006, their Fig.~9), with the VLA. The low resolution ($\sim $ 78\arcsec ) \ion{H}{1} map shows (unequal) double-peaked \ion{H}{1} emission, almost horizontal (PA$_{\rm HI} \approx $ 10$^\circ$) disk morphology, which is coincident with the optical position angle (PA$_{\rm opt}\approx $ 10$^\circ$). There are faint extensions to the North and South. 
%{\it \ion{H}{1}/opt = 3; Type n/a}

{\bf Sextans B (also known as UGC 5373 or DDO 070)} From Ott et al. (2012, their Fig.~10), from the VLA ANGST. The high resolution ($\sim $ 6\arcsec) \ion{H}{1} map shows a complex, quasi-spherical, spiral-like morphology, with \ion{H}{1} clumps and holes. The regions of more enhanced \ion{H}{1} emission appear to lie mainly above the optical disk and on the Southeast side, in an inverted "C-shape", so that the internal \ion{H}{1} morphology is roughly aligned with the optical disk (PA$_{\rm HI} \sim $ PA$_{\rm opt} \approx $ 25$^\circ$). The direction of the velocity gradient (PA$_{\rm VG}\approx $ 130$^\circ$) is virtually perpendicular to the (internal) \ion{H}{1} and optical morphology. 
%{\it \ion{H}{1}/opt = 3; Type A/B}

{\bf Sextans A (also known as UGCA 205 or DDO 075)} From Ott et al. (2012; their Fig.~13), from the VLA ANGST. The high resolution ($\sim $ 6\arcsec) \ion{H}{1} image shows an elongated (PA$_{\rm HI} \approx $ 135$^\circ$) disk-like structure, with \ion{H}{1} clumps and gaps. The \ion{H}{1} morphology is approximately perpendicular to the optical axis (PA$_{\rm opt}\approx $ 45$^\circ$). The \ion{H}{1} is enhanced on either side of the optical disk, roughly positionally coincident with the brightest star formation knots, with opposite extensions to the North and South. The velocity gradient direction (PA$_{\rm VG} \approx $ 150$^\circ$) is slightly misaligned with the \ion{H}{1} morphology, but roughly perpendicular to the optical disk. 
%{\it \ion{H}{1}/opt = 2; Type B}

{\bf UGC 6456 (also known as VII Zw 403)} From Thuan, Hibbard \& L\'evrier (2004; their Fig.~17, 18 and 20), and Lelli, Verheijan \& Fraternali (2014a, b; their Fig.~1, C.12, and Fig.~1, 2), from the VLA. The intermediate resolution (20\arcsec~x 20\arcsec, 21.2\arcsec~x 14.5\arcsec, 17.2\arcsec~x 11\arcsec~and 15\arcsec~x 15\arcsec) \ion{H}{1} maps show double-peaked disk emission (PA$_{\rm HI} \approx $ 75$^\circ$), extended to the Northwest, South and West. The \ion{H}{1} emission does not align with the main optical emission (PA$_{\rm opt}\approx $ 115$^\circ$), but does align with the faint extended emission (PA$_{\rm opt}\approx $ 75$^\circ$). The main \ion{H}{1} peak coincides, not with the center of the optical disk, but with the brightest star forming regions to the South in the optical image. The velocity field is disturbed, and can be modeled by a combination of rotation and radial motions. The kinematical axis (PA$_{\rm kin}\approx $ 90$^\circ$) is vertical, with the rotation occurring along the \ion{H}{1} major axis. 
%{\it \ion{H}{1}/opt $\sim$ 2; Type B}

{\bf SBS 1129+576} From Ekta, Pustilnik \& Chengalur (2003; their Fig.~1, 3 and 6), with the GMRT. The low resolution (42\arcsec~x 40\arcsec) \ion{H}{1} map shows the XMP target, along with its companion galaxy, SBS 1129+577, $D \sim $ 27~kpc to the North, and the dwarf galaxy SDSS J113227.68+572142 to the Southeast. The image signals the interaction of the SBS 1129+567/577 pair, by way of a \ion{H}{1} bridge connecting both galaxies, and a common \ion{H}{1} envelope. There is also a small tail of emission to the Northeast and Southwest of the XMP. At this resolution, the position angle of the \ion{H}{1} emission associated with the XMP target is PA$_{\rm } \approx $ 75$^\circ$ , which is roughly coincident with the direction of the velocity gradient (PA$_{\rm VG}\approx $ 70$^\circ$). The velocity field is fairly ordered (solid-body rotation) in the central region and also in the bridge, consistent with a prograde merger. The high resolution (6\arcsec~x 14\arcsec) map shows a Northwest-Southeast disk (PA$_{\rm HI} \approx $ 70$^\circ$), in good agreement with the optical axis (PA$_{\rm opt} \approx $ 75$^\circ$). The disk displays small extensions to the South, East and West, similar to what is seen at lower resolution. The position of the \ion{H}{1} peak encompasses both the optical disk center and the bright knot to the Southeast. At high resolution (8\arcsec~x 7\arcsec), the \ion{H}{1} peak is resolved into two high density regions, each positionally coincident with the optical disk center and the bright star formation knot. Similarly to the low resolution image, extended emission to the South, East and West is detected. The somewhat curved tips of the \ion{H}{1} disk may suggest a warp. The \ion{H}{1} position angle on small-scales (PA$_{\rm HI} \approx $ 70$^\circ$) is roughly in agreement with that on intermediate-large scales, with the direction of the velocity gradient and with the optical disk. 
%{\it \ion{H}{1}/opt $\sim$ 2; Type A/B}

{\bf UGCA 292 (also known as NGC 128)} From Young et al. (2003; their Fig.~2), with the VLA, and Ott et al. (2012; their Fig.~22), from the VLA ANGST. Intermediate resolution (17.7\arcsec~x 17.4\arcsec~and 14.2\arcsec~x 13.9\arcsec) \ion{H}{1} maps show triple-peaked, "boomerang-shaped" \ion{H}{1} emission along an almost horizontal axis (PA$_{\rm HI} \approx $ 175$^\circ$). The internal \ion{H}{1} inverted "V-shape" structure mimics what is observed in the optical (PA$_{\rm opt}\approx $ 165$^\circ$). Indeed, the \ion{H}{1} peaks are positionally coincident with the ends and center of the optical disk, where bright star formation knots can be found. The vertical gradient direction (PA$_{\rm VG}\approx $ 145$^\circ$) is misaligned by about 20$^\circ$ from the overall \ion{H}{1} and optical emission, but is roughly aligned with the Southwest "boomerang" optical/\ion{H}{1} arm. The velocity field displays an asymmetry in the periphery. The high resolution ($\sim $ 6\arcsec) \ion{H}{1} spatial distribution and velocity field is a scaled-down version of the lower resolution observations. 
%{\it \ion{H}{1}/opt $\sim$ 3; Type A/B}

{\bf GR 8 (also known as DDO 155, UGC 8091 or VIII Zw 222)} From Young et al. (2003; their Fig.~3) and Begum \& Chengalur (2003; their Fig.~2, 3 and 4), with the VLA, Begum et al. (2006; their Fig.~9, 12 and 13,), with the GMRT, and Ott et al. (2012; their Fig.~23), from the VLA ANGST. At low and intermediate resolution (30\arcsec~x 30\arcsec~and 25\arcsec~x 25\arcsec), the overall \ion{H}{1} emission is triple-peaked, enclosed in a "cross-shaped" halo (PA$_{\rm HI} \approx $ 165$^\circ$), where the central triple-peak emission is "V-shaped". The kinematical axis (PA$_{\rm kin}\approx $ 80$^\circ$) is misaligned with both the \ion{H}{1} and the optical disk emission (PA$_{\rm opt}\approx $ 130$^\circ$). The velocity field can be modeled by a combination of circular plus radial motions of the gas, showing distortions in the outskirts. Intermediate resolution (18.6\arcsec~x 18.1\arcsec~and 14.8\arcsec~x 14.6\arcsec) and high resolution ($\sim $ 6\arcsec~and 4\arcsec~x 3\arcsec) \ion{H}{1} images mimic the triple-peaked "V-shaped" structure (PA$_{\rm HI} \approx $ 165$^\circ$) seen at lower resolution. These peaks are slightly misaligned relative to the position of the three brighter star-forming regions in the optical disk. 
%{\it \ion{H}{1}/opt $\sim$ 2; Type A/B}

{\bf SBS 1415+437} From Lelli, Verheijan \& Fraternali (2014a, b; their Fig. C.18, and Fig.~1, 2), with the VLA. The intermediate resolution (20\arcsec~x 20\arcsec) \ion{H}{1} map shows a single-peaked Northeast-Southwest (PA$_{\rm HI} \approx $ 115$^\circ$) disk, extended to the North, and with smaller extensions to the South, East and West. The \ion{H}{1} peak is positionally coincident, not with the center of the optical disk, but with the Southern bright star formation knot that heads the cometary structure, while the cometary tail is mirrored in the Northeastern extension. High resolution (10\arcsec~x 10\arcsec) \ion{H}{1} observations show a similar morphology to the intermediate resolution image, but where the extended South, East and West emission is now resolved into a web of small tails and filaments. The \ion{H}{1} emission at intermediate-high resolution scales is roughly aligned with the optical disk (PA$_{\rm opt} \approx $ 120$^\circ$) and the kinematical axis (PA$_{\rm kin} \approx $ 115$^\circ$). The velocity field is ordered (rotation) only in the center, with the outskirts displaying lopsidedness. 
%{\it \ion{H}{1}/opt $\sim$ 2; Type A}

{\bf Sag Dig} From Begum et al. (2006; their Fig.~9), with the GMRT, and Becarri et al. (2014; their Fig.~12 and 13), with the VLA. The low resolution ($\sim $ 67\arcsec) \ion{H}{1} map shows double-peaked quasi-spherical ring-like emission, with an offcenter \ion{H}{1} depression ($\Delta D \lesssim $ 1\arcmin). The brightest \ion{H}{1} peak roughly coincides with the bright star formation knots to the East, seen on the optical image (PA$_{\rm opt} \approx $ 170$^\circ$). At intermediate resolution (30\arcsec~x 30\arcsec), the ring-like structure becomes more apparent, with multiple \ion{H}{1} clumps and a large offcenter \ion{H}{1} hole. The velocity field displays no sign of ordered rotation. Becarri et al (2014) hypothesized that this source could be the result of a merger with another dwarf. 
%{\it \ion{H}{1}/opt = 6; Type B/C}

{\bf J2104-0035} From Ekta, Chengalur \& Pustilnik (2008; their Fig.~5 and 7), with the GMRT. The low resolution (29\arcsec~x 25\arcsec) \ion{H}{1} map shows a single-peaked inverted "triangular" morphology, with several small extensions to the North and a tail to the Southeast. The \ion{H}{1} peak is not coincident with the bright star formation knot at the Northern edge of the cometary structure, but instead is coincident with the Southern edge of the optical disk. The overall \ion{H}{1} morphology (PA$_{\rm HI} \approx $ 90$^\circ$) is slightly misaligned from the optical structure (PA$_{\rm opt} \approx $ 110$^\circ$), but aligned with the velocity gradient (PA$_{\rm VG}\approx $ 90$^\circ$). The velocity field displays ordered (rotation) in the central part, but a North-South asymmetry is present in the outskirts. The intermediate and high resolution (17\arcsec~x 11\arcsec) map reveals "Y-shaped" emission, with extensions to the South, North and Northwest. The overall \ion{H}{1} morphology is better aligned with the optical structure. In the highest resolution (7\arcsec~x 5\arcsec) image, the "Y-shape" is more pronounced, cradling the bright star formation knot in the North. There is extended emission to the South and a clump of emission to the Northwest, mimicking the intermediate resolution image. 
%{\it \ion{H}{1}/opt = 5; Type A/B}

% high density regions
% discontinuous/lopsided

\end{appendix}

\end{document}